\documentclass[twocolumn,showpacs,amsmath,amssymb]{revtex4}
\usepackage{graphicx}
\usepackage{hyperref}

\begin{document}

\title{Evolutionary Dynamics for Persistent Cooperation in Structured Populations}

\author{Yan Li, Xinsheng Liu}
\affiliation{ Institute of Nano Science, State Key Laboratory of Mechanics and Control of Mechanical Structures, Key Laboratory for Intelligent Nano Materials and Devices of the Ministry of Education, and Department of Mathematics, Nanjing University of Aeronautics and Astronautics, Nanjing 210016, China}

\author{Jens Christian Claussen}
\email{j.claussen@jacobs-university.de}
\affiliation{Jacobs University Bremen, Computational Systems Biology Lab, Research 2, Campus Ring 1, D-28759 Bremen, Germany}

\author{Wanlin Guo}

\affiliation{Institute of Nano Science, State Key Laboratory of Mechanics and Control of Mechanical Structures, Key Laboratory for Intelligent Nano Materials and Devices of the Ministry of Education, and Department of Mathematics, Nanjing University of Aeronautics and Astronautics, Nanjing 210016, China}

\date{14.\ November 2014; revised 13. May 2015; accepted 18. May 2015}
\mbox{}\\[-5ex]  \mbox{}\hspace*{0.75\textwidth}\fbox{To appear in \href{http://pre.aps.org}{Physical Review E}}\\
\begin{abstract}
The emergence and maintenance of cooperative behavior is a fascinating topic in evolutionary biology and social science. 
The public goods game (PGG) is a paradigm for exploring cooperative behavior. In PGG, the total resulting payoff is divided equally among all participants. This feature still leads to the dominance of defection without substantially magnifying the public good by a multiplying factor. Much effort has been made to explain the evolution of cooperative strategies, including a recent model in which only a portion of the total benefit is shared by all the players through introducing a new strategy named persistent cooperation. A persistent cooperator is a contributor who is willing to pay a second cost to retrieve the remaining portion of the payoff contributed by themselves. In a previous study, this model was analyzed in the framework of well-mixed populations. This paper focuses on discussing the persistent cooperation in lattice-structured populations. The evolutionary dynamics of the structured populations consisting of three types of competing players (pure cooperators, defectors and persistent cooperators) are revealed by theoretical analysis and numerical simulations. In particular, the approximate expressions of fixation probabilities for strategies are derived on one-dimensional lattices. The phase diagrams of stationary states, the evolution of frequencies and spatial patterns for strategies are illustrated on both one-dimensional and square lattices by simulations. 
Our results are consistent with the general observation that, at least in most situations, a structured population facilitates the evolution of cooperation.
Specifically, here we find that the existence of persistent cooperators greatly suppresses the spreading of defectors under more relaxed conditions in structured populations compared to that obtained in well-mixed population.  
\end{abstract}
\pacs{87.23.Cc,89.65.-s,02.50.Le,87.23.Ge}
\maketitle

\section{Introduction}
How to understand the evolution of cooperative behavior has interested researchers for decades. Evolutionary game theory
\cite{weibull93,nowak06book,claussen08}
provides a powerful framework and effective methods for studying the emergence of cooperation. Among many game models, the public goods game (PGG) 
\cite{traulsen05,binmore94,kagel97,santos08}
-- which can be viewed 
as a generalization of the Prisoner’s Dilemma with pairwise interactions \cite{nowak93,nowaksigmund93} --
serves as a paradigm 
for discovering the mechanism of cooperative behavior both in theory and experimental settings
\cite{brandt03}.
 In a PGG, players are required to decide simultaneously whether or not to invest in the common pool, i.e., to cooperate or to defect. All participants share the public resource equally irrespective of their individual contribution. Obviously, individuals obtain a better one-shot game payoff if they invest nothing into the public goods. A selfish population which overexploits the public goods will finally be trapped in the “tragedy of the commons” \cite{hardin68}. Though the existence of selfishness and competitiveness in human society, field studies and experiments have verified the fact that humans are willing to cooperate if the conditions are appropriate \cite{sigmund10book}. Many mechanisms were put forward to enhance cooperation in social dilemmas. For example, 
social diversity \cite{santos08} and
voluntary participation in PGG turned out to be a simple but viable means to overcome this difficulty 
\cite{hauert02}. Reward (rewarding cooperators at personal cost and benefiting cooperators) and punishment (punishing defectors by paying a cost and diminishing the defectors) are both effective ways to promote and sustain cooperation 
which,  on the other hand, raises the problem of so called second free riders who exploit the additional contributions of rewarding players and punishers
\cite{szabo02,sigmund01,sigmund07}.
Punishment is a mechanism that can facilitate altruism also in animal societies \cite{clutton95}. 
However, almost all the previous published works based on PGG assumed that the total payoff 
is
shared equally among all the members engaged in the game whatever strategies they selected. It is no surprise that the traditional way of payoff distribution used in PGG cannot provide positive incentive for players to cooperate if no other mechanisms exist.
\par
Recently, Liu and Guo proposed a new model consisting of three competing strategies, i.e.\ persistent cooperation, pure cooperation and defection \cite{liu10}, aiming to address the enigma: why cooperators would choose to punish defectors at personal cost. In their model, cooperators invest in the public goods at personal cost while defectors contribute nothing. Only a proportion of the total payoff is shared among all members, which is different from the traditional way used in PGG. The contributors who are willing to pay a second cost to retrieve the remaining part of the profit invested by themselves are called persistent cooperators ($PC$). The model is discussed in well-mixed populations where all the individuals meet each other at random with equal probability. The persistent cooperation strategy can evolve under somewhat strict conditions. However in practical situations the populations are often structural, and individuals have little chance to interact with anyone randomly and they can only contact with their neighbors in a certain range. Many researchers have devoted 
efforts to the study of 
evolutionary games in spatially structured populations \cite{brandt03,szabo02,traulsen04,lieberman05,ohtsuki06,szolnoki10},
whereby space can both facilitate and suppress the evolution of cooperation \cite{nowak92,hauert04}.
\par
In this paper we discuss the model of persistent cooperation 
in 
lattices, 
which -- albeit being the simplest type of a 
structured population -- 
provide an even field for competing strategies where already the possibility of
network reciprocity is given \cite{nowak92}.
For this reason, lattices
 enjoy significant popularity in game theoretical models \cite{szabofath07,nakamaru97},
and,
regardless of their difference to the actual social networks
\cite{wassermann94},
 they give us a very helpful entry point for studying the impact of the structure on the evolution of cooperation
\cite{perc13}.
\par
   The remainder of this the paper is organized as follows. In section \ref{sec2},
 the game models on lattices are introduced in detail, and then simulations and theoretical analysis of the evolutionary dynamics of the strategies are presented for one-dimensional lattice in section \ref{sec3}.
The phase diagrams of stationary states, the evolution of frequencies and spatial patterns for strategies on square lattice are provided in section \ref{sec4}. In section \ref{sec5}, a  summary is made based on the analysis and computer simulations in previous sections. 

\section{Models of Persistent Cooperation on Lattices \label{sec2}}
\par
    The PGG is arranged on a lattice with periodic boundary conditions to remove the edge effect. Each player on site $x$ initially is designated either as a cooperator ($C$), a persistent cooperator ($PC$) or a defector ($D$) with equal probability. Players play the game with $n$ immediate neighbors around them, i.e., each individual is the focal player of the group 
 he (or she)
belongs to. Using standardized parameters here, both $PC$-players and $C$-players invest an amount 1 into the common pool, while defectors contribute nothing. The sum of all contributions is multiplied by a factor $r>$1, reflecting the synergic effects of cooperation. Let $P_{x,n_{PC},n_C}$ be the payoff that an $x-$player earns, if he interacts with $n_{PC}$ persistent cooperators, $n_C$ cooperators and   $n_D=n-n_{PC}-n_C$ defectors.
\par
The overall payoff 
per game round 
is divided in two stages: in the first stage, only a fraction $s (0<s<1)$ of the resulted benefit is shared equally among $n+1$ participators irrespective of their strategies. 
\par
In the second stage, the remaining proportion of the total income $(1-s)r(n_{PC}+n_C)$  is divided into two parts: 
One part  $(1-s)rn_{PC}$ contributed by the persistent cooperators will be retrieved by themselves with persistent efforts at a second personal cost $dn_D/(n+1)$, where  $n_D/(n+1)$  is the proportion of defectors and $d>0$. Thus in this stage each persistent cooperator gets the payoff $(1-s)r-dn_D/(n+1)$. 
In contrast,
the pure cooperators are unwilling to bear 
any
additional cost to 
receive a 
 part of deserved payback $(1-s)rn_C$, so this part of payoff is again
(not equally) 
shared 
among cooperators and defectors. Each defector gets $(1-s)rn_C/(n+1)$, a cooperator reaps $[(1-s)rn_C-(1-s)rn_Cn_D/(n+1)]/n_C$. Here we assume that the $PC$-players 
have no intention to
share the $(1-s)$ part of the payoff contributed by the cooperators. 
\par
 Therefore, the total payoff for each player engaged in the game are:
\begin{eqnarray}
P_{C,n_{PC},n_C}
&=&\frac{r(n_{PC}+n_C+1)}{n+1}-1
\label{eq7}
\\
\label{eq8}
P_{D,n_{PC},n_C}
&=&\frac{rn_C+srn_{PC}}{n+1}
\\
P_{{PC},n_{PC},n_C}
&=&
\frac{r(n_c+n_{PC}+1)}{n+1} - 1 
+\frac{[(1-s)r-d]n_D}{n+1}
\nonumber \\&& \label{eq9}
\end{eqnarray}
\par
Note
 that the second cost  $dn_D/(n+1)$ is directly proportional to the frequency of the defectors $n_D/(n+1)$. 
In a biological setting, in a population 
 of lions and hyenas, lions pay a first cost to kill buffalos. However, they have to pay a second cost to fight hyenas for keeping the prey from being robbed 
\cite{mills91,clutton95}.
The value of the second cost depends on the number of hyenas, and determines the future action of lions: persisting or giving up. The parameter $d$ in $ dn_D/(n+1)$  represents the degree of the second cost (persistent cooperating) and it is not necessarily identical to 1. The strategy $PC$ will be less and less favored by selection if the value of $d$ becomes larger and larger. 
With the increase of $s$, the defectors get more and more benefit from the PGG without any cost, they fare better than the cooperators and the persistent cooperators, thus, all the players in the group are prone to become “free riders”. Such transitions perhaps lead to tragic consequences as in normal PGG. 
In the following sections,
we 
investigate
 the evolution of the competing strategies in dependence upon $s$ and $d$ for typical synergic factor $r$.
\par
All simulations are initialized with a uniform spatial distribution of the three strategies,
and we perform a random sequential update where
each Monte Carlo step (MCs) 
is defined as follows:
\begin{enumerate}
\item
A randomly selected player $x$ plays the PGG with $n$ neighbors in a group composed of $n_{PC}$  persistent cooperators, $n_C$ cooperators and $n_D=n-n_{PC}-n_C$ defectors, and obtains his payoff.
\item
One of the $n$ nearest interacting partners of $x$ is selected at random, denoted by $y$,
and obtains his payoff.
\item 
Player $x$ imitates the strategy of $y$ with probability 
%
$Q=1/(1+\exp(w(P_{x,n_{PC},n_C}-P_{y,n_{PC},n_C})))$.
\end{enumerate}
 In this “pairwise comparison” update
\cite{szabotoke98},
$w$  corresponds to the inverse temperature in statistical physics, acting as selection intensity. 
This choice of $Q$ also resembles the ``logit rule'' employed to model human choice behavior
\cite{mcfadden73,mcfadden76}.
As $w \ll 1$ (weak selection, the payoff has little effect on the strategy selection) 
\cite{traulsen07}, $Q\to 0.5$, selection becomes neutral. If $w\to\infty $ (strong selection), 
we arrive at
$ Q\to1$, or $Q\to 0$, depending solely on the sign of the payoff difference, hence the
"pairwise comparison" updating rule becomes deterministic,
 indicating that an individual always adopts his neighbor with higher income and refuses to imitate one with lower payoff 
\cite{hofbauer03}. In our simulation, without losing generality, we set $w=2$, implying that better performing individuals are readily to be imitated. 
By the finite choice of $w$,
 we never exclude the situation that a player adopts the strategy of an individual performing worse than him.  
\par

\par
It is known that the system size can influence the dynamics to a large degree, so we have confirmed our model in different linear system sizes, e.g., 1000,  4000, and 10000 for one-dimensional lattice, and the equilibrium needs up to $7\times10^4$ full Monte Carlo steps. 
For the square lattice, we perform simulations on two different system sizes $100\times100$ and $400\times400$, reaching the equilibrium needs up to $10^5$ full Monte Carlo steps. The resulting phase diagrams have similar distributions in the phase planes. These simulations indicate that our results are robust in even larger systems.


\section{Persistent Cooperation on a one-dimensional Lattice \label{sec3}}
\subsection{Simulation Results}
\begin{figure}
  \centering
  \includegraphics[height=0.22\textheight]{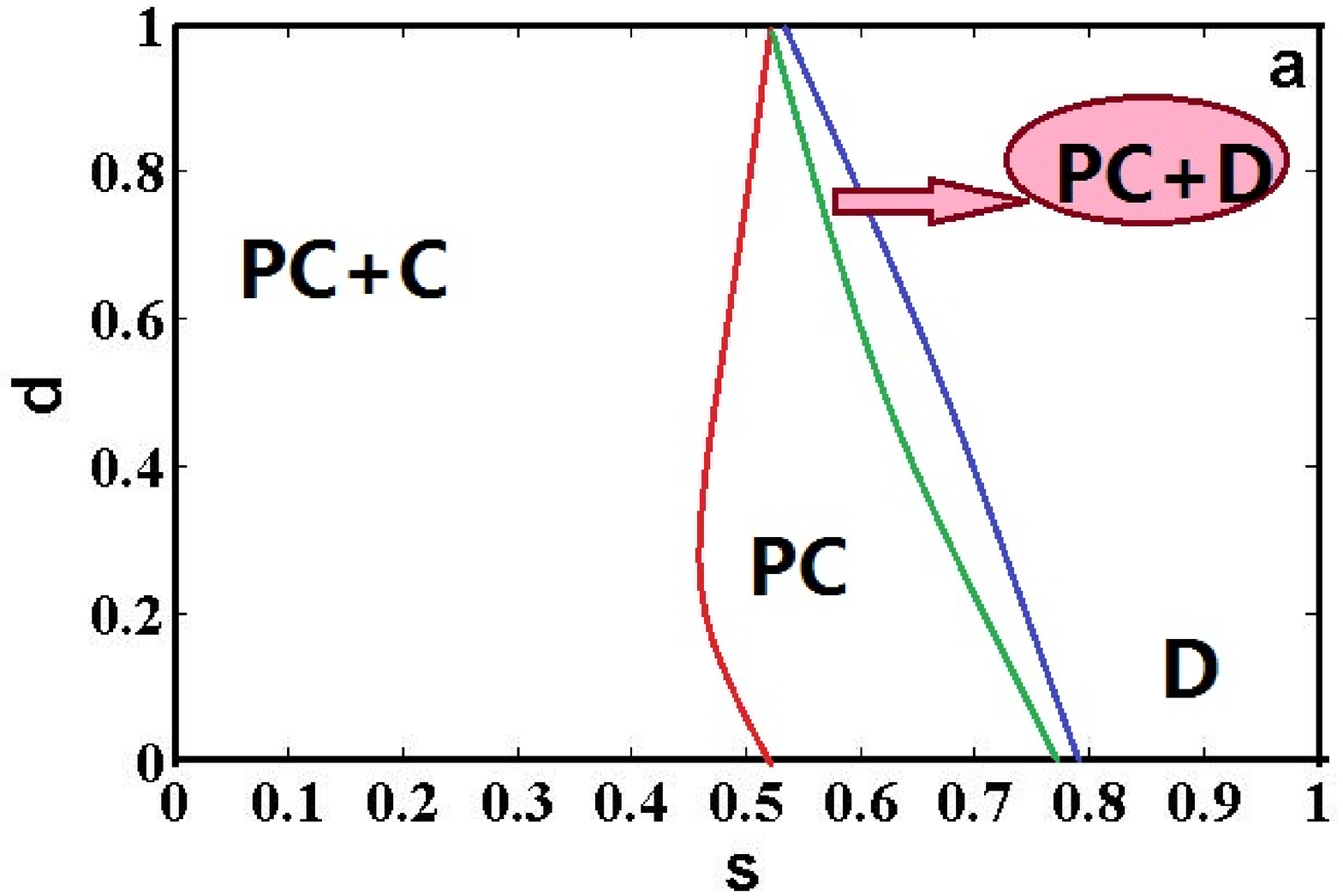} 
  \includegraphics[height=0.21\textheight]{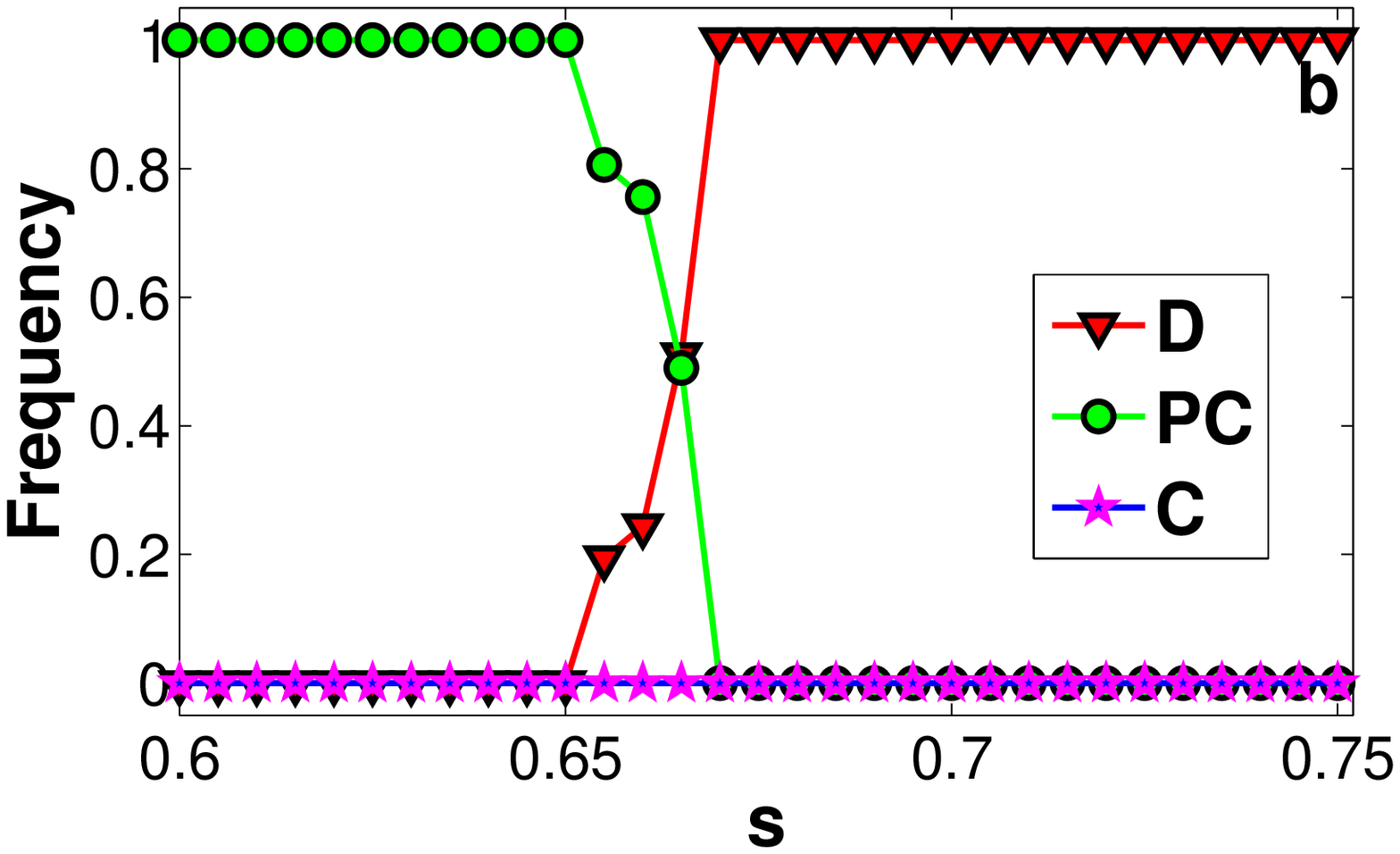} \\
 \includegraphics[height=0.23\textheight]{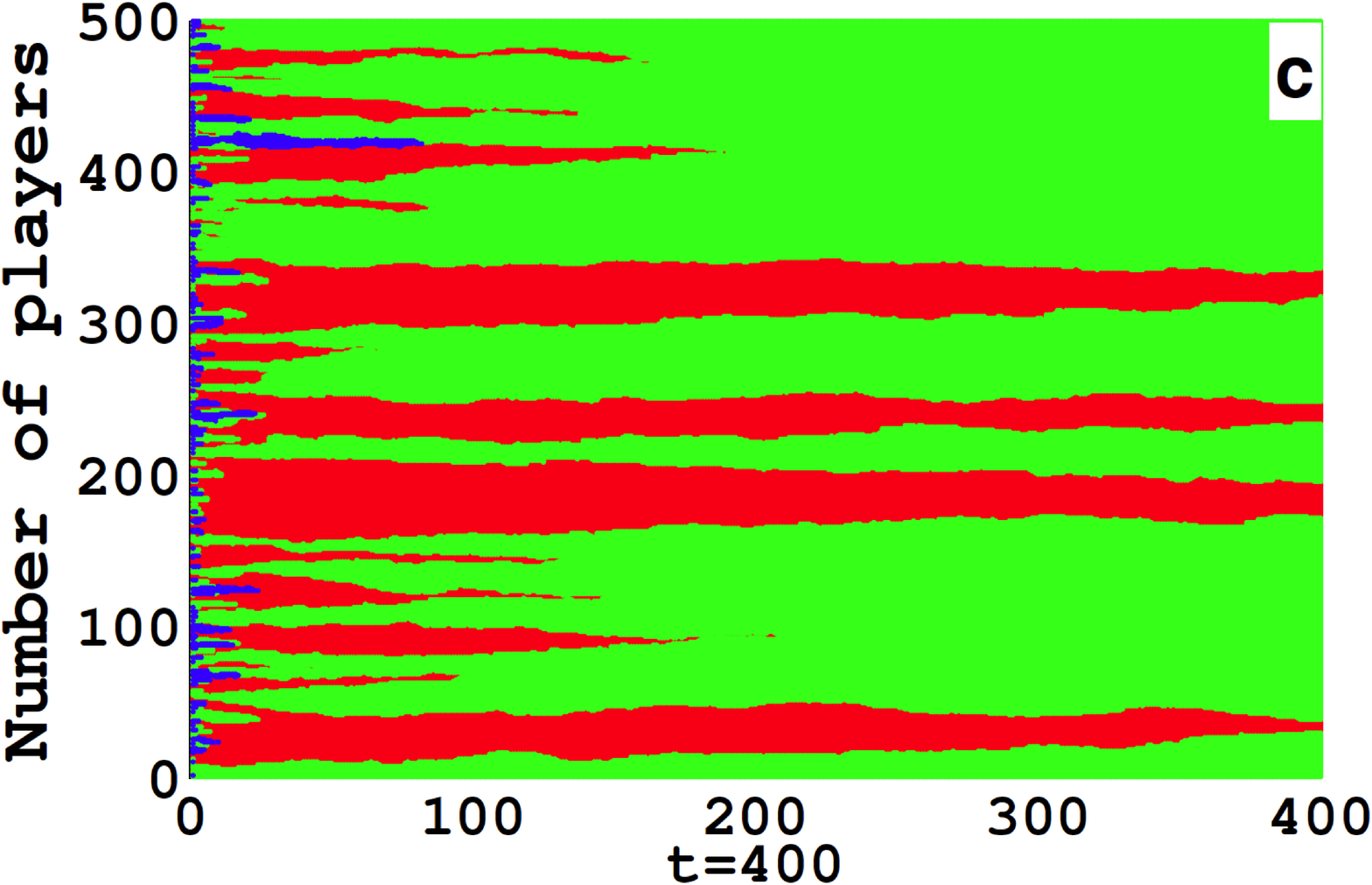}
  \includegraphics[height=0.185\textheight]{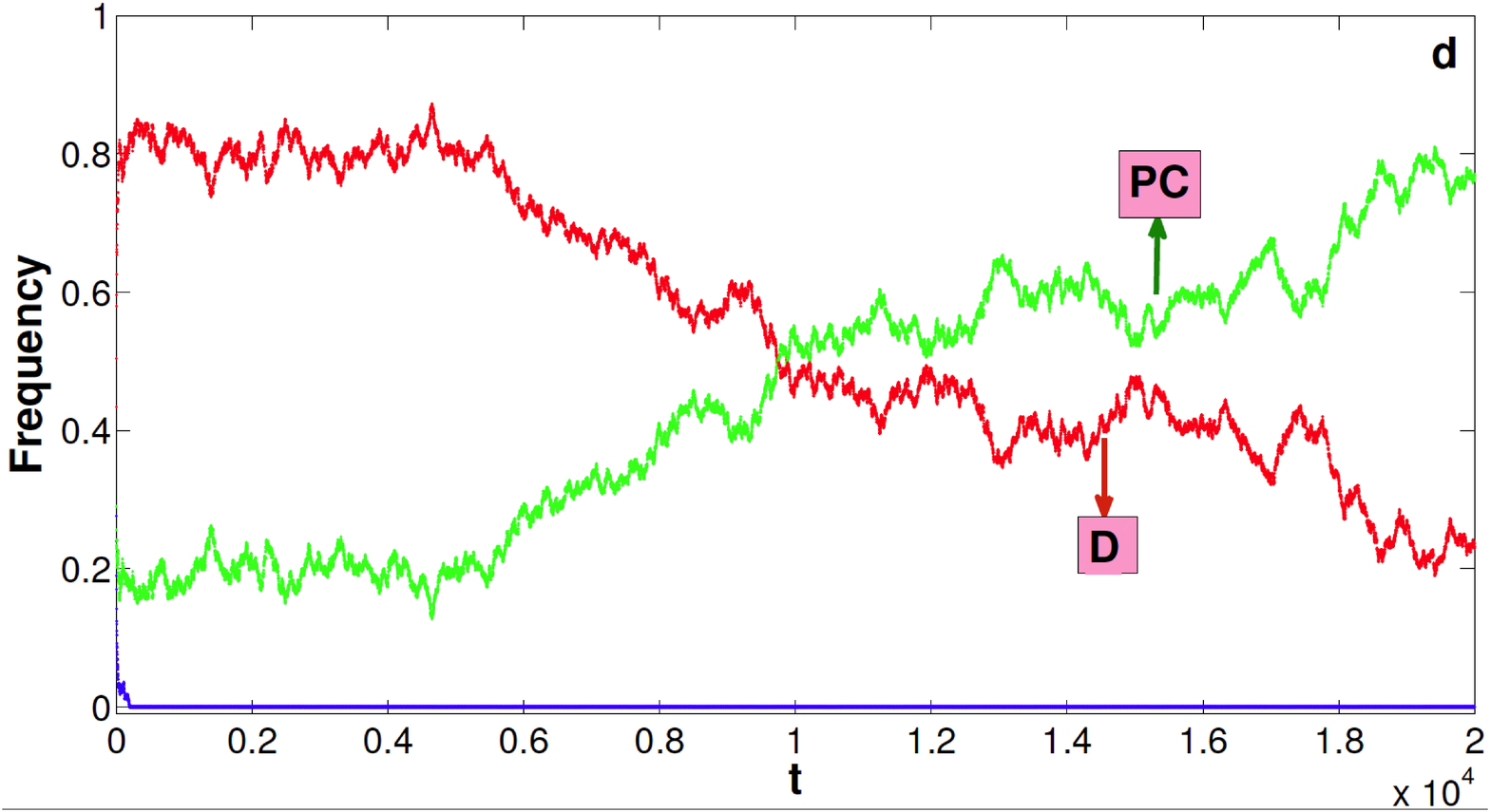}
  \caption{(Color online)   (a) Full $s-d$ phase diagram obtained for $r=2, w=2$. Different phases are denoted by the symbols of strategies that survive finally in the equilibrium state. Solid lines indicate continuous transitions between different states. (b) A representative cross-section of (a) at $d=0.4$ illustrating the frequencies of $C, D$ and $PC$ depending on s. (c) is a spatial pattern of the model indicating how the three strategies evolves within 400 full MCs, where $s=0.55, d=0.4$. (d) The changes of the three strategies with time as $s=0.66, d=0.4$. \label{fig1}}
\end{figure}

We perform numerical simulations on a one-dimensional lattice of size 500, where each focal individual has $n=2$ adjacent neighbors and on its left and right side, respectively. Fig.~\ref{fig1}a illustrates the full phase diagram based on $s$ and $d$. The synergic factor $r$ is 2. It can be seen that the phase plane is partitioned into four regions denoted by the symbols of strategies that survive in the final equilibrium state, $PC+C$ (the coexistence of strategy $PC$ and $C$), $PC, PC+D$ (the coexistence of strategy $PC$ and $D$) and $D$. The regions of the defectors and the cooperators ($PC$ and $C$) are almost equal. If $s<0.5$, i.e., less than half of the total profit is shared among all the players, although the defectors can exploit the pure cooperators, their incomes are cut down by the $PC$-players, so the defectors vanish in this region due to their low fitness. With the increasing of $s$, defectors get more and more payoff, whose income are higher than that of the cooperators, hence, the $C$-players give way to the defectors. The battle of territory remains between the persistent cooperators and the defectors. The narrow region denotes the coexistence of strategy $PC$ and $D$.
\par
A representative cross-section of (a) at $d=0.4$ illustrating the frequencies of $C, D$ and $PC$ in dependence on $s$ is shown in Fig.~\ref{fig1}b. If $0.5\leq s\leq 0.65$, the battle among the three strategies ends up with the dominance of the persistent cooperators. When $0.65<s<0.67$, the defectors come up, whose frequency increases from zero to one quickly. 
While $s\geq 0.67$, the defectors win the game, the persistent cooperators disappear completely. 
\par
Fig.~\ref{fig1}c is a spatial pattern of the model indicating how the three strategies evolve within 400 full MCs. Starting from a random initial design, the cooperators vanish soon in the competition due to their lower income. So the battle mostly occurs only between the persistent cooperators and the defectors. Players with the same strategy form compact clusters quickly to fight against their opponents. We find that the variations of the evolutionary dynamics are not remarkable between two adjacent full Monte Carlo steps. The main reason is that there are countable boundaries in one-dimensional lattice-structured population. The transitions of strategies only take place at the boundaries of these tight clusters, so most of the individuals have little opportunity to interact with their opponents once the clusters have established. The frequencies of two adjacent interacting pairs of players with different strategies (e.g., $PC-D$ and $C-D$) approach zero
since the interface density decays with time with a scaling that depends on
the type of update \cite{bennaim96}.
The scaling of characteristic domain length during a similar three-species coarsening process has been elucidated in
\cite{frachebourg96,bennaim96redner} for the case of cyclic coevolution
and it can be anticipated that similar scaling laws hold also in the non-cyclic case. 
\par
 Fig.~\ref{fig1}d elucidates the dynamics of the three strategy frequencies when $s=0.66$ and $d=0.4$. The frequency of strategy $C$ decreases to zero soon, which is quite different from the feature exhibited in well-mixed population. In well-mixed populations, the time for strategy $C$ involved in the contest is much longer than that in lattice-structured populations.

\subsection{Fixation Probability}
\par
In well-mixed population, if $(1-s)r<d+1$, i.e.\ $s>1- (d+1)/r$ ($s>0.44$ for $r=2.5, d=0.4$), defection is the only evolutionary stable strategy and can repel the invasion of rare persistent cooperators
\cite{liu10}. So it is quite difficult for persistent cooperation to establish itself in a population dominated by defection initially. However, in a one-dimensional lattice model, numerical simulations show that a persistent cooperator can invade a population dominated by defectors successfully and become fixed when $s=0.45$ $(>0.44)$. Motivated by this observation, we are to find the condition for successful invasion and fixation of $PC (D)$ in a one-dimensional lattice-structured population.

\begin{figure}
\centering
  \includegraphics[height=0.12\textheight]{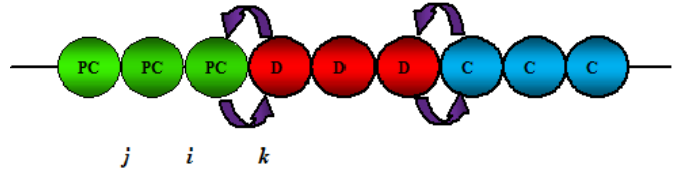}
\caption{(Color online.)  Players with different strategies are aligned on one dimensional lattice, transitions only occur at the boundary sites between two runs of players with different strategies, and then the focal individual will be replaced by one of its nearest neighbors. \label{fig2}}
\end{figure}

\par
As illustrated in Fig.~\ref{fig2}, the strategy transitions only occur at the boundary sites between two clusters of players with different strategies, which follow a random-walk, and the focal individual will be replaced by one of his nearest neighbors according to pair comparison. 
\par
Let $f_n(t)$ be the probability that a cluster consisting of $n$ players with the same strategy (named $n$-cluster) at time $t$ survives and takes over the whole population in the end. $\alpha_n$ is the transition probability from a $n$-cluster to a ($n+1$)-cluster in a unit time, while $\beta_n$ is the transition probability from a $n$-cluster to a ($n-1$)-cluster. The $n$-cluster remains the same with probability $1- \alpha_n - \beta_n$. The basic equations obtained by birth-death stochastic process are:
\begin{equation}\label{eq10}
\left\{
\begin{aligned}
f_0(t)=&0\\
f_n(t)=&f_n(t)(1- \alpha_n - \beta_n)+f_{n+1}(t)\alpha_n\\
&+f_{n-1}(t)\beta_n,   1\leq n\leq{N-1}\\
f_N(t)=&1\\
\end{aligned}
\right.
\end{equation}
\par
A direct calculation \cite{taylor04} shows that the fixation probability, which is defined as the probability that a mutant invading a population of $N-1$ resident individuals will produce a lineage which takes over the whole population 
\cite{nowak04,karlintaylor75}, is given by
\begin{equation}\label{eq11}
f_1(t)=\frac{1}{1+\sum_{n=1}^{N-1}{\prod_{j=1}^n{\beta_j/\alpha_j}}}
\end{equation}\\
\par
The most critical step here is to find the transition probabilities $\alpha_n$ and $\beta_n$, $n$=1,2,\dots,$N$. Firstly, let $x_{PC}$  and $x_D$  be the global frequencies of strategy $PC$ and $D$ on one-dimensional lattice, respectively. Secondly, we regard $j-i-k$ as a strategy-triplet. Each focal $i$-player has $n=2$ adjacent neighbors ($j$-player and $k$-player) on his left and right side, as illustrated in Fig.~\ref{fig2}. By the frequency of the triplet denoted by $x_{i-j-k}$ , $i, j, k\in \{C, D, PC\}$, the dynamics of  $PC$ and $D$   are expressed by the following differential equations: 

\begin{eqnarray}\label{eq12}
\dot{x}_{PC}&=& -Q_{PC,1,0}x_{D-PC-PC} - Q_{PC,1,1}x_{PC-PC-C}\nonumber\\
&&-Q_{PC,0,2}x_{C-PC-C}-Q_{PC,0,0}x_{D-PC-D}\nonumber\\
&&- Q_{PC,0,1}x_{D-PC-C}(PC\to D)\nonumber\\
&&- Q_{PC,0,1}x_{D-PC-C}(PC\to C)\\
&&+Q_{D,1,0}x_{PC-D-D}+Q_{D,1,1}x_{PC-D-C}\nonumber\\
&&+Q_{D,2,0}x_{PC-D-PC}+Q_{C,1,0}x_{PC-C-D}\nonumber\\
&&+Q_{C,1,1}x_{PC-C-C}+Q_{C,2,0}x_{PC-C-PC}\nonumber
\end{eqnarray}

\begin{eqnarray}\label{eq13}
\dot{x}_{D}&=& -Q_{D,1,0}x_{PC-D-D} - Q_{D,0,1}x_{D-D-C}\nonumber\\
&&-Q_{D,2,0}x_{PC-D-PC}-Q_{D,0,2}x_{C-D-C}\nonumber\\
&&- Q_{D,1,1}x_{PC-D-C}(D\to PC)\nonumber\\
&&- Q_{D,1,1}x_{PC-D-C}(D\to C)\\
&&+Q_{PC,0,1}x_{D-PC-C}+Q_{PC,1,0}x_{D-PC-PC}\nonumber\\
&&+Q_{PC,0,0}x_{D-PC-D}+Q_{C,1,0}x_{PC-C-D}\nonumber\\
&&+Q_{C,0,1}x_{D-C-C}+Q_{C,0,0}x_{D-C-D}\nonumber
\end{eqnarray}
\par
The first term on the right hand side in (\ref{eq10}) is the rate that a focal $PC$-player surrounded by a $D$-player and a $PC$-player on his left and right side switches to be a defector in a unit time. It is a product of the imitation probability $Q_{PC,1,0}=1/(1+\exp(w(P_{PC,1,0}-P_{D,1,0})))$ (the focal $PC$-player interacts with a defector and a cooperator on its left and right side respectively, and then imitates the strategy of the defector) and the frequency of the triplet $D-PC-PC$. If and only if the $PC$-player imitates the strategy of the $D$-player,  the frequency of $PC$  can change. 
The negative sign takes care of the
negative change of the frequency of $PC$. Because the triplet $PC-PC-D$ has the same function as $D-PC-PC$, there is a factor 2 for the frequency of $D-PC-PC$, but it is canceled by another factor 0.5, the probability that each individual is selected on both sides of the $PC$-player. Similar explanations hold for other frequencies of strategy-triplets. Notably, though the two expressions $ Q_{PC,0,1}x_{D-PC-C}(PC\to D)$ and $ Q_{PC,0,1}x_{D-PC-C}(PC\to C)$ in the second row in (\ref{eq10}) have the common term $ Q_{PC,0,1}x_{D-PC-C}$, they represent different cases. One case is that the $PC$-player imitates the strategy of the $D$-player, which leads to the reduction of the number of $PC$-players by one. The other case is that the $PC$-player imitates the strategy of the $C$-player, which also leads to the reduction of the number of $PC$-players by one. We use arrows $PC\to C$ and $PC\to D$ to distinguish the two cases. Similarly the terms  $Q_{D,1,0}x_{PC-D-D}$, $Q_{D,1,1}x_{PC-D-C}$ and $Q_{D,2,0}x_{PC-D-PC}$ ($Q_{C,1,0}x_{PC-C-D}$, $Q_{C,1,1}x_{PC-C-C}$ and $Q_{C,2,0}x_{PC-C-PC}$) are the transitions from a $D$-player (C-player) to a $PC$-player, which result in positive changes of the frequency of strategy $PC$.
\par
Lastly, we suppose that the configurations are uniformly distributed in space. From the definition of the process, the transition probabilities for strategy $PC$ can be given by the following expressions:
\begin{equation}\label{eq14}
 \alpha_n^{PC}=
\begin{cases}
\frac{1}{4}[Q_{D,1,0}+Q_{D,1,1}+Q_{C,1,0}\\+Q_{C,1,1}]\triangleq\alpha^{PC},& n=1,2,\dots,N-2\\
 \frac{1}{2}(Q_{D,2,0}+Q_{C,2,0})\triangleq\alpha_{N-1}^{PC},&n=N-1
\end{cases}
\end{equation}
\begin{equation}\label{eq15}
 \beta_n^{PC}=
\left\{
  \begin{array}{ll}
    \frac{1}{4}[Q_{PC,0,2}+Q_{PC,0,0}\\+Q_{PC,0,1}(PC\to C)\\+Q_{PC,0,1}(PC\to D)]\triangleq\beta_1^{PC}, & \hbox{$n=1$} \\
    \frac{1}{2}(Q_{PC,1,0}+Q_{PC,1,1})\triangleq\beta^{PC}, & \hbox{$n=2,\dots,N-1$.}
  \end{array}
\right.
\end{equation}
The correspondent transition probabilities for strategy $D$ are:
\begin{equation}\label{eq16}
 \alpha_n^{D}=
\begin{cases}
\frac{1}{4}[Q_{PC,0,1}+Q_{PC,1,0}+Q_{C,0,1}\\+Q_{C,1,0}]\triangleq\alpha^{D},& n=1,2,\dots,N-2\\
 \frac{1}{2}(Q_{PC,0,0}+Q_{C,0,0})\triangleq\alpha_{N-1}^{D},&n=N-1
\end{cases}
\end{equation}
\begin{equation}\label{eq17}
 \beta_n^{D}=
\left\{
  \begin{array}{ll}
    \frac{1}{4}[Q_{D,0,2}+Q_{D,2,0}\\+Q_{D,1,1}(D\to PC)\\+Q_{D,1,1}(D\to C)]\triangleq\beta_1^{D}, & \hbox{$n=1$} \\
    \frac{1}{2}(Q_{D,1,0}+Q_{D,0,1})\triangleq\beta^{D}, & \hbox{$n=2,\dots,N-1$.}
  \end{array}
\right.
\end{equation}
Therefore, according to (\ref{eq13}), the fixation probability of $PC (D)$ is
\begin{eqnarray}\label{eq18}
 f^i_1(t)\!\!&=&\!\!\frac{1}{1\!+\!\frac{\beta_1^i}{\alpha^i}\!+\!\sum_{n=2}^{N-2}{
    \frac{\beta^i_1}{\alpha^i}(\frac{\beta^i}{\alpha^i})^{n-1}}\!+\!
    \frac{\beta_1^i}{\alpha^i}\frac{\beta^i}{\alpha^i_{N-1}}(
    \frac{\beta^i}{\alpha^i})^{N-3}}   ~~~~~~~ \\
 &&i\in \{PC,D\}\nonumber
\end{eqnarray}
If the number of individuals in this lattice-structured population is large enough, the fixation probability may be approximated by
\begin{equation}\label{eq19}
 f_1^{i}(t)=
\begin{cases}
\frac{\alpha^i-\beta^i}{\alpha^i-\beta^i+\beta^i_1},& \alpha^i>\beta^i\\
 0,&\alpha^i\leq\beta^i
\end{cases}
\end{equation}
\par
Similar procedures can be used for finding the fixation probability of strategy $C$. In particular, if the population is only composed of persistent cooperators and defectors, the expressions of fixation probabilities of $PC$ and $D$ are the same as in equation (\ref{eq19}), where
\begin{equation*}
 \alpha_n^{PC}=
\begin{cases}
Q_{D,1,0},& n=1,2,\dots,N-2\\
 Q_{D,2,0}&n=N-1
\end{cases}
\end{equation*}

\begin{equation}\label{eq20}
 \beta_n^{PC}=
\left\{
  \begin{array}{ll}
    Q_{PC,0,0}, & \hbox{$n=1$} \\
    Q_{PC,1,0} & \hbox{$n=2,\dots,N-1$.}
  \end{array}
\right.
\end{equation}

\begin{equation*}
 \alpha_n^{D}=
\begin{cases}
Q_{PC,1,0},& n=1,2,\dots,N-2\\
Q_{PC,0,0},&n=N-1
\end{cases}
\end{equation*}

\begin{equation}\label{eq21}
 \beta_n^{D}=
\left\{
  \begin{array}{ll}
   Q_{D,2,0}, & \hbox{$n=1$} \\
   Q_{D,1,0}, & \hbox{$n=2,\dots,N-1$.}
  \end{array}
\right.
\end{equation}
\par
\begin{figure}
\centering
  \includegraphics[height=0.21\textheight]{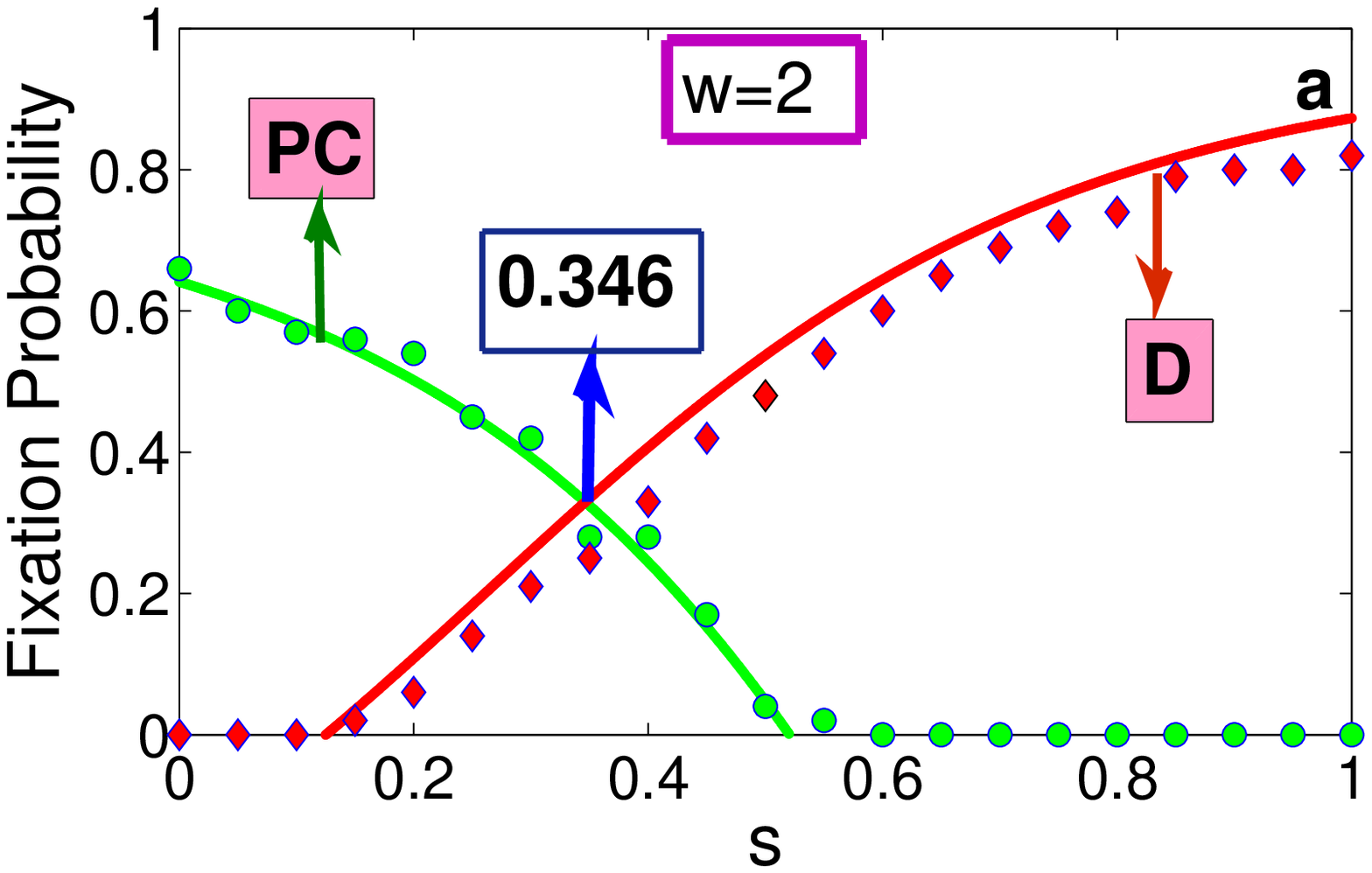} 
  \includegraphics[height=0.21\textheight]{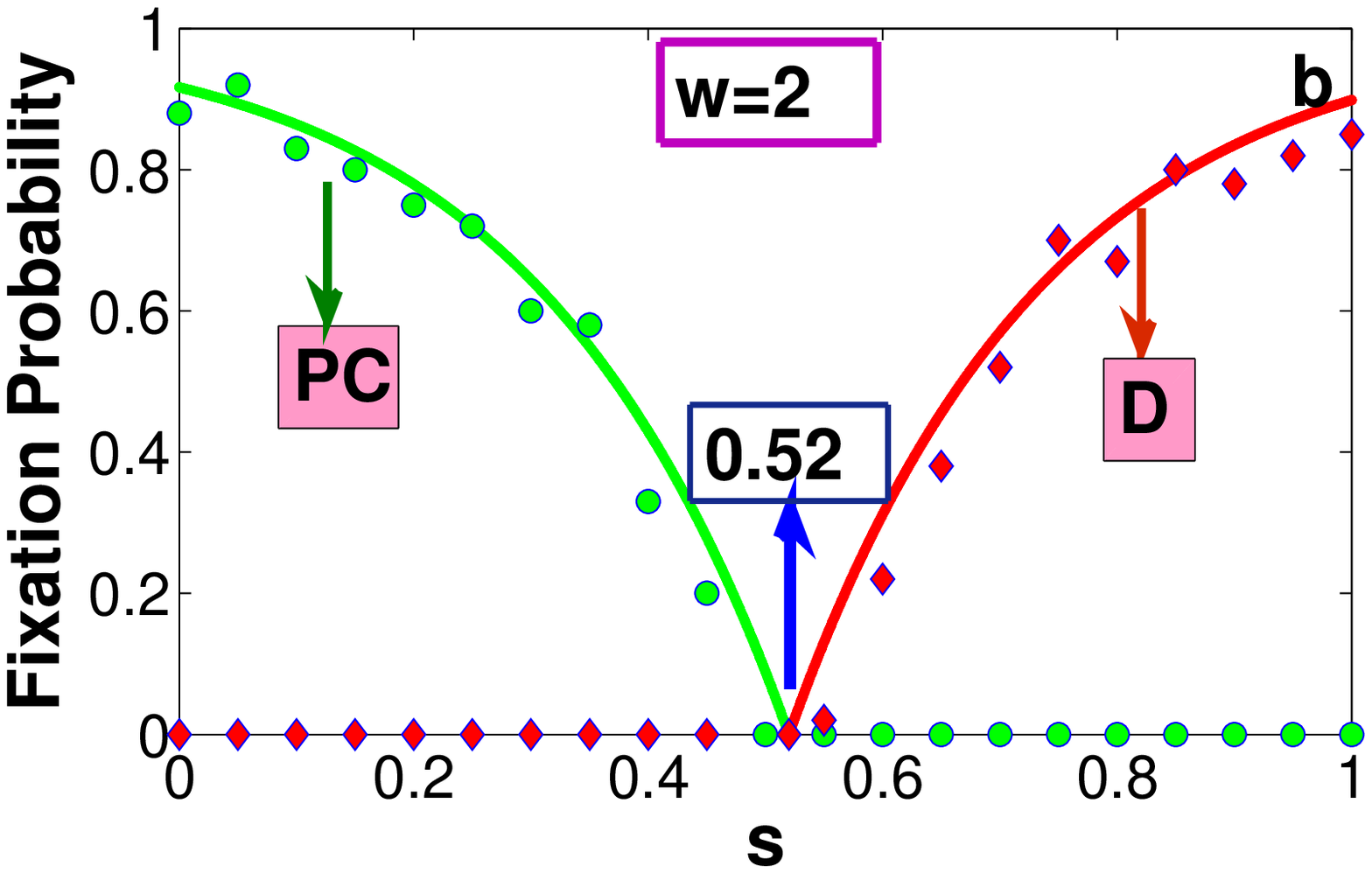} \\
  \includegraphics[height=0.21\textheight]{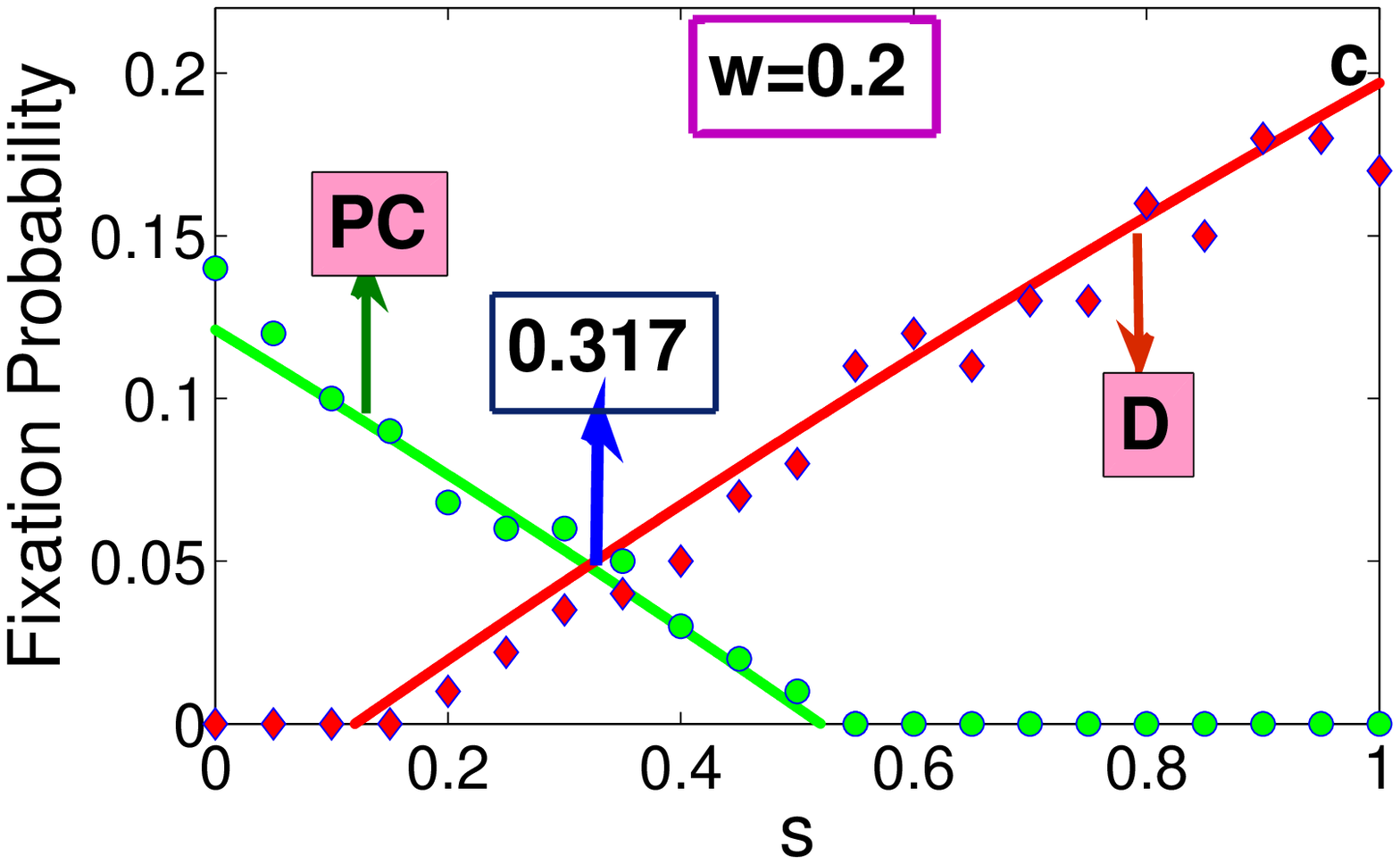}
  \includegraphics[height=0.21\textheight]{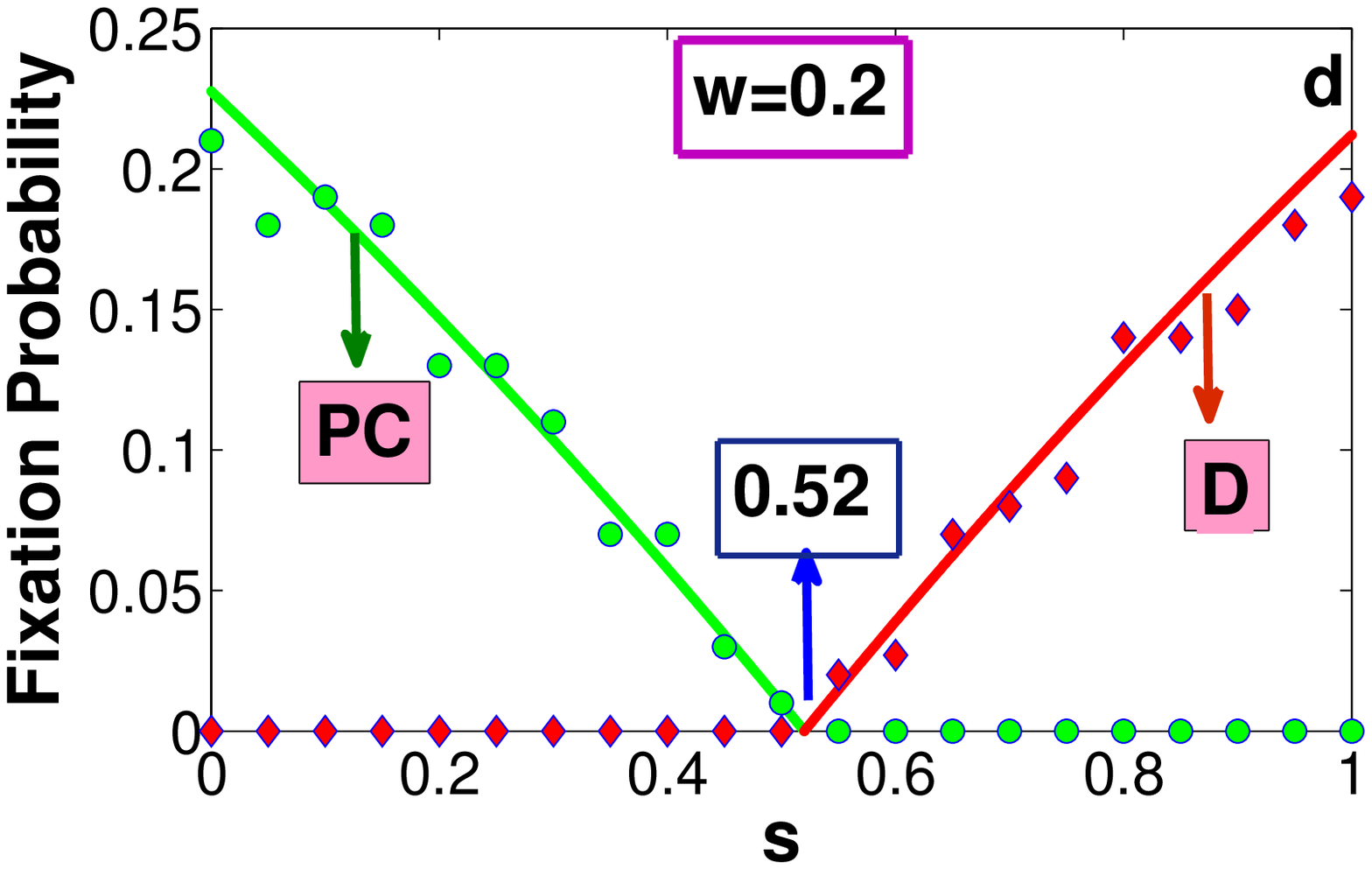}
 \caption{(Color online)  (a) and (c): The fixation probabilities for strategy $PC$ (green line) and $D$ (red line) in a one-dimensional lattice-structured population with $C, D$ and $PC$. (b) and (d): The fixation probabilities for strategy $PC$ (green line) and $D$ (red line) in one-dimensional lattice-structured population with only $D$ and $PC$. Parameters are set as $r=2.5, d=0.4$ for all (a) to (d). Simulations are performed on $500\times1$ lattice for the fixation probabilities of $PC$ (green solid dots) and $D$ (red solid diamonds), respectively. \label{fig3}}
\end{figure}

Now, to make a direct comparison with the results obtained in well-mixed population, we analyze the transition probabilities in (\ref{eq14}) and (\ref{eq15}) under the limit of weak selection by Taylor expansion under which we have: 
\begin{eqnarray}
\alpha^{PC}&\approx& \frac{1}{2}+\frac{w}{8}[(1-s)r-1-2k/3]\nonumber\\
\beta^{PC}&\approx& \frac{1}{2}+\frac{w}{8}[k/3+1-(1-s)r] \label{eq22}
\end{eqnarray}
and
\begin{eqnarray}
\alpha^{D}&\approx& \frac{1}{2}+\frac{w}{8}[k/3+2-(1-s)r]\nonumber\\
\beta^{D}&\approx& \frac{1}{2}+\frac{w}{8}[(1-s)r-2k/3-2] \label{eq23}
\end{eqnarray}
We distinguish two cases:
\begin{description}
\item[(1)] If $1+\frac{d}{2}<r(1-s)<\frac{3+d}{2}$, i.e.\ $1-\frac{3+d}{2r}<s<1-\frac{2+d}{2r}$  ($0.317<s<0.52$, as can be observed in Fig.~\ref{fig3}c), though the fixation probability of $D$ is larger than that of PC in this range, we still have $\alpha^{PC}>\beta^{PC} $ which follows that a single $PC$-player has a positive probability to invade and become fixed in a one-dimensional lattice-structured population occupied by $C$-players and $D$-players. However, in this case the $PC$-players can not be fixated in  a well-mixed population \cite{liu10}.
\item[(2)] If $r(1-s)>\frac{3+d}{2}$ i.e. $s<\frac{3+d}{2r}$ ($s<0.317$), which is equivalent to the conditions $\alpha^{PC}>\beta^{D} $ and $\beta^{PC}>\beta^{D} $ we reach the conclusion $f_1^{PC}>f_1^{D}$. In this case the $PC$-players are rather advantageous. A single $PC$-player is easier to invade, survive and fix in the population than a single $D$ individual.
\end{description}

\par
The fixation probabilities for $PC$ (green line) and $D$ (red line) on a one-dimensional lattice predicted by equation (19) are shown in Fig.~\ref{fig3}, respectively. Parameters are $r=2.5$ and $d=0.4$ for all panels (a) to (d). Figs.~\ref{fig3}a$+$c show that in a one-dimensional lattice-structured population with three strategies $PC,$ $C$ and $D$, a single $PC$-player can invade successfully, reproduce inchmeal and become fixed with a positive probability if $s<0.52$. 
For
$s>0.52$, a single $PC$ fails to invade in the population, and cannot escape from the bad luck of going extinct. Figures \ref{fig3}a$+$c present a little difference because of the different values of $w$. Notably, a $PC$ invader can be fixed easier than a $D$ player if $s<0.346$ for $w=2$ (0.317 for $w=0.2$). When $s>0.346$ (0.317 for $w=0.2$), the opposite case occurs. 
Fig.~\ref{fig3}b ($w=2$) and Fig.~\ref{fig3}d ($w=0.2$) illustrate the fixation probabilities of $PC$ and $D$ in a population of only persistent cooperators and defectors, respectively. $s=0.52$ is the threshold for $PC$ and $D$ being evolution stable strategy, as labeled in Figs.~\ref{fig3}b$+$d. While in well-mixed population composed of only $PC$-players and $D$-players, persistent cooperators dominate defectors if $s<1-(1+d)/r =0.44(<0.52)$ for $r=2.5, d=0.4$. Due to the existence of cooperators, the range of $s$ for the fixation of $D$ is somewhat larger than that of $PC$.
\par
Numerical simulation results on a one-dimensional lattice of finite size 500 are presented in Fig.~\ref{fig3} by green solid dots for $PC$ and red diamonds for $D$. Just for convenience to construct the model to obtain the fixation probabilities, we simply assume that the strategy-triplets are uniformly located on the lattice, but it is more realistic to take into account the aggregation degree of the same strategy and the density of the population on the lattice model. When performing numerical simulations, we let the strategies be randomly located on the one-dimensional lattice. As shown in Fig.~\ref{fig3}, the simulation results are well consistent with the theoretical results obtained by formula (\ref{eq19}).
\par
\begin{figure}
\centering
  \includegraphics[height=0.22\textheight]{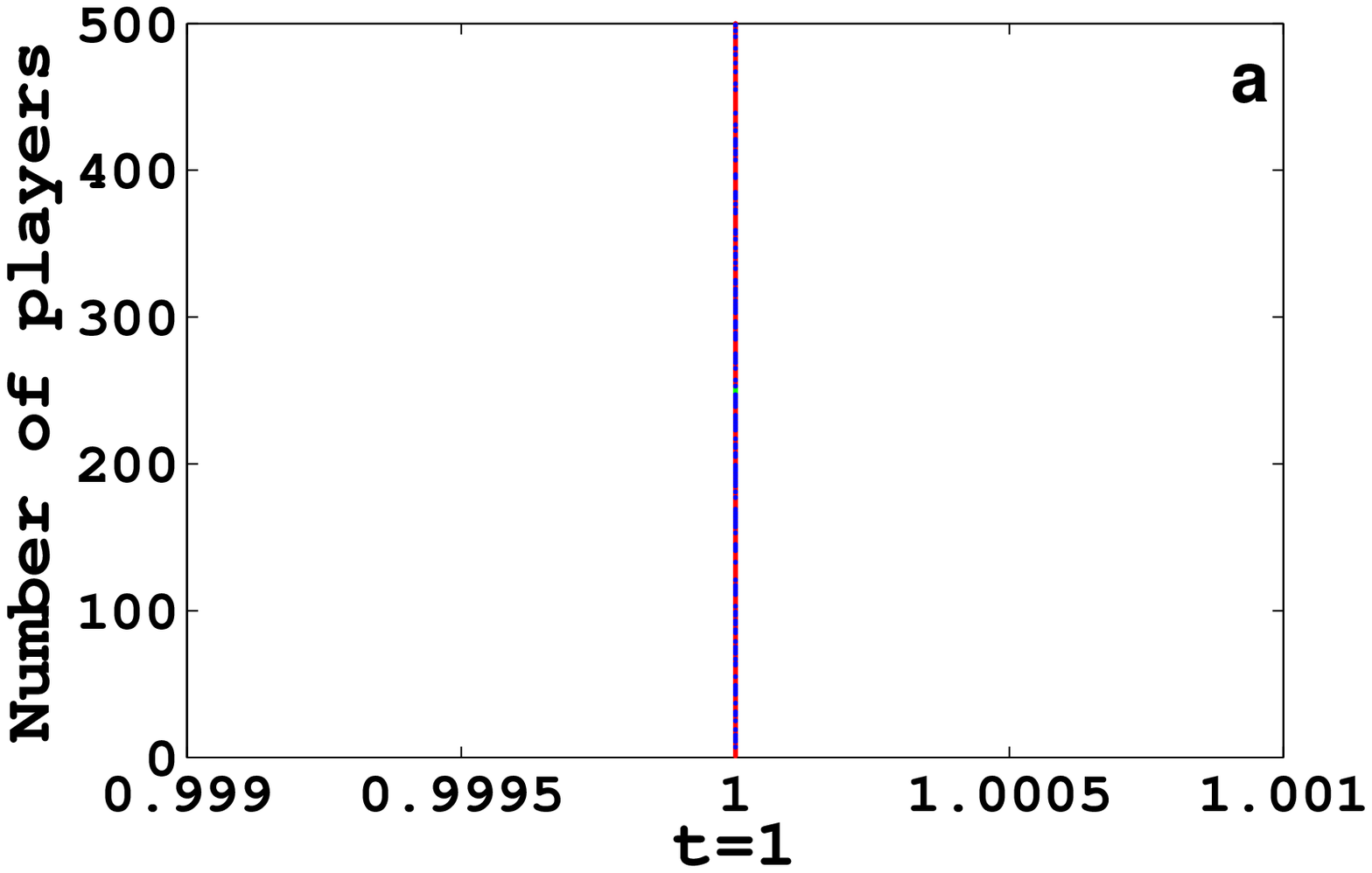} 
 \includegraphics[height=0.21\textheight]{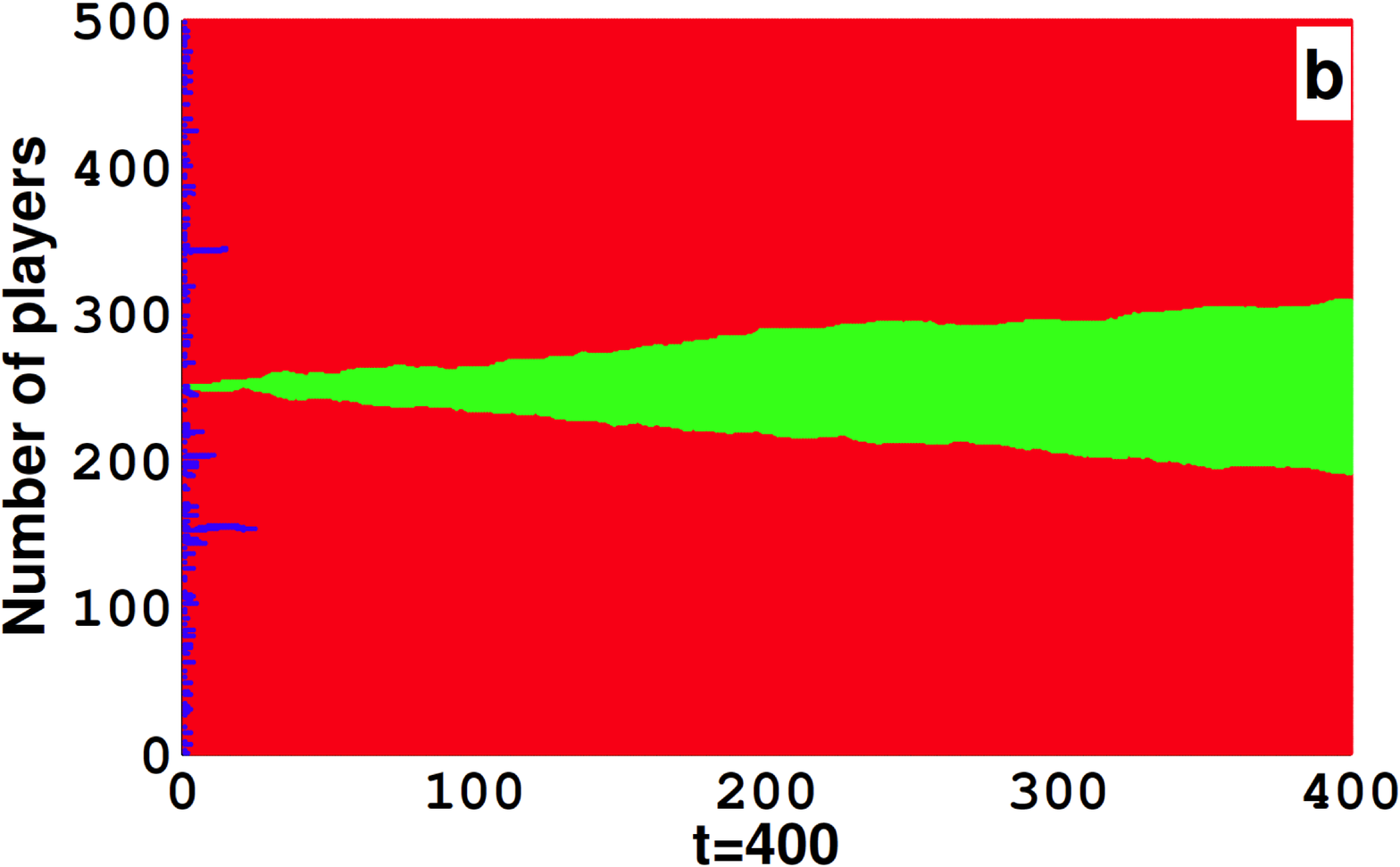} \\
  \includegraphics[height=0.225\textheight]{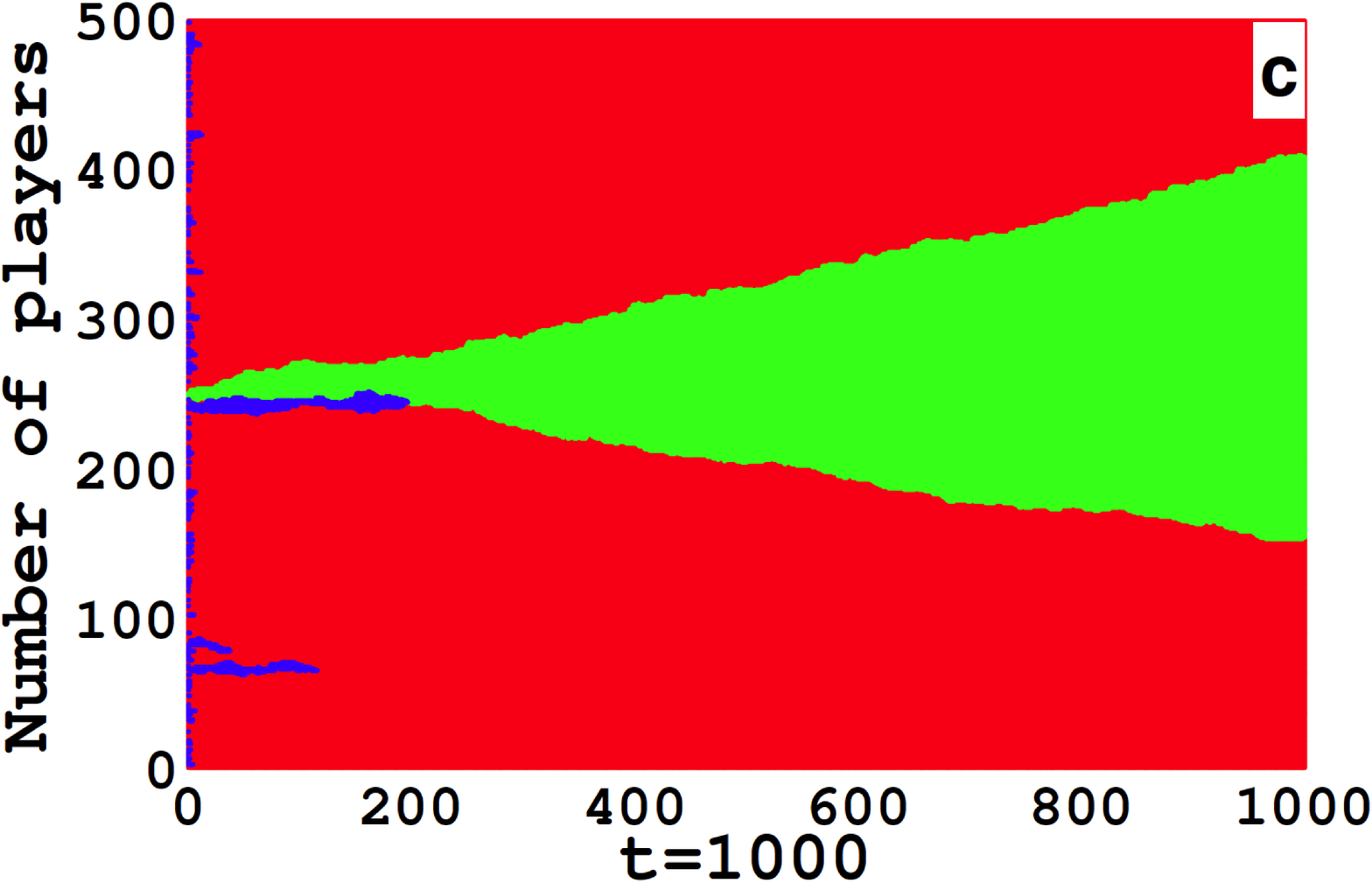}
  \caption{(Color online) A single $PC$-player invades successfully, reproduces inchmeal and becomes fixed in a 
500$\times$1
lattice-structured population. (a) - (c) are characteristic snapshots of the evolution dynamics over time for $PC$. The parameters are $r=2.5, w=2, s=0.45$ and $d=0.4$. Colors (blue, green and red) show the distribution of cooperators, persistent cooperators and defectors. \label{fig4}}
\end{figure}

As shown in Fig.~\ref{fig4}, a single $PC$-player invades the one-dimensional lattice-structured population of $C$ and $D$. Since the cooperators are always exploited by the defectors without the “protection” of the $PC$-players, the number of the $C$-players becomes less and less. The pure cooperators disappear in a short time. 
But single $PC$-players, surrounded by defectors who exploit pure cooperators, 
can proliferate
 and grow offspring forming a compact cluster drawing back the
$(1-s)$
proportion of the payoff contributed by themselves from the defectors.
The defectors are defeated in the end.

\section{Persistent Cooperation on a Square Lattice \label{sec4}}
\par
In this section, we shall consider persistent cooperation on square lattice. After the three strategies are assigned uniformly at random on a square lattice, a randomly selected individual plays the game with $n=4$ neighbors
 (von Neumann neighborhood) 
around him, i.e. each player is the focal one of the group. 
\par
In a square lattice-structured population with only cooperators and defectors in previous studies \cite{szolnoki09}, with similar values of the parameters in our model, cooperators survive only if $r >3.74$ and crowd out defectors completely for $r >5.49$. These results can be taken as benchmarks for evaluating the impact of persistent cooperation on the 
evolutionary 
dynamics of cooperation in structured populations. In this section, we focus on discussing how different combinations of $s$ and $d$ affecting the evolution of the three strategies based on two different synergic factor $r=2$ and 3.5, representing for low and high synergetic effects of cooperation, respectively.
\begin{figure}
\centering
  \includegraphics[height=0.22\textheight]{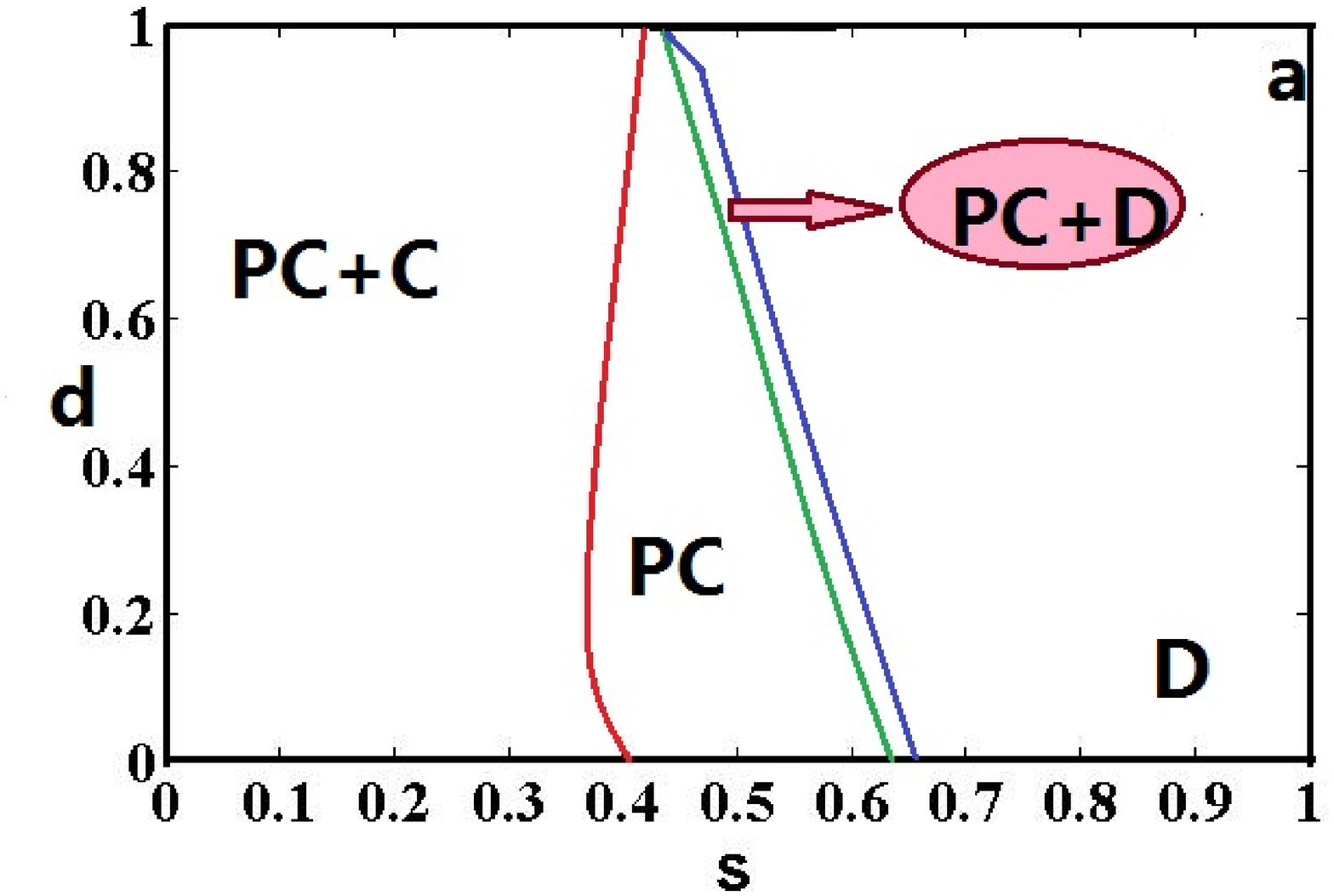} 
  \includegraphics[height=0.22\textheight]{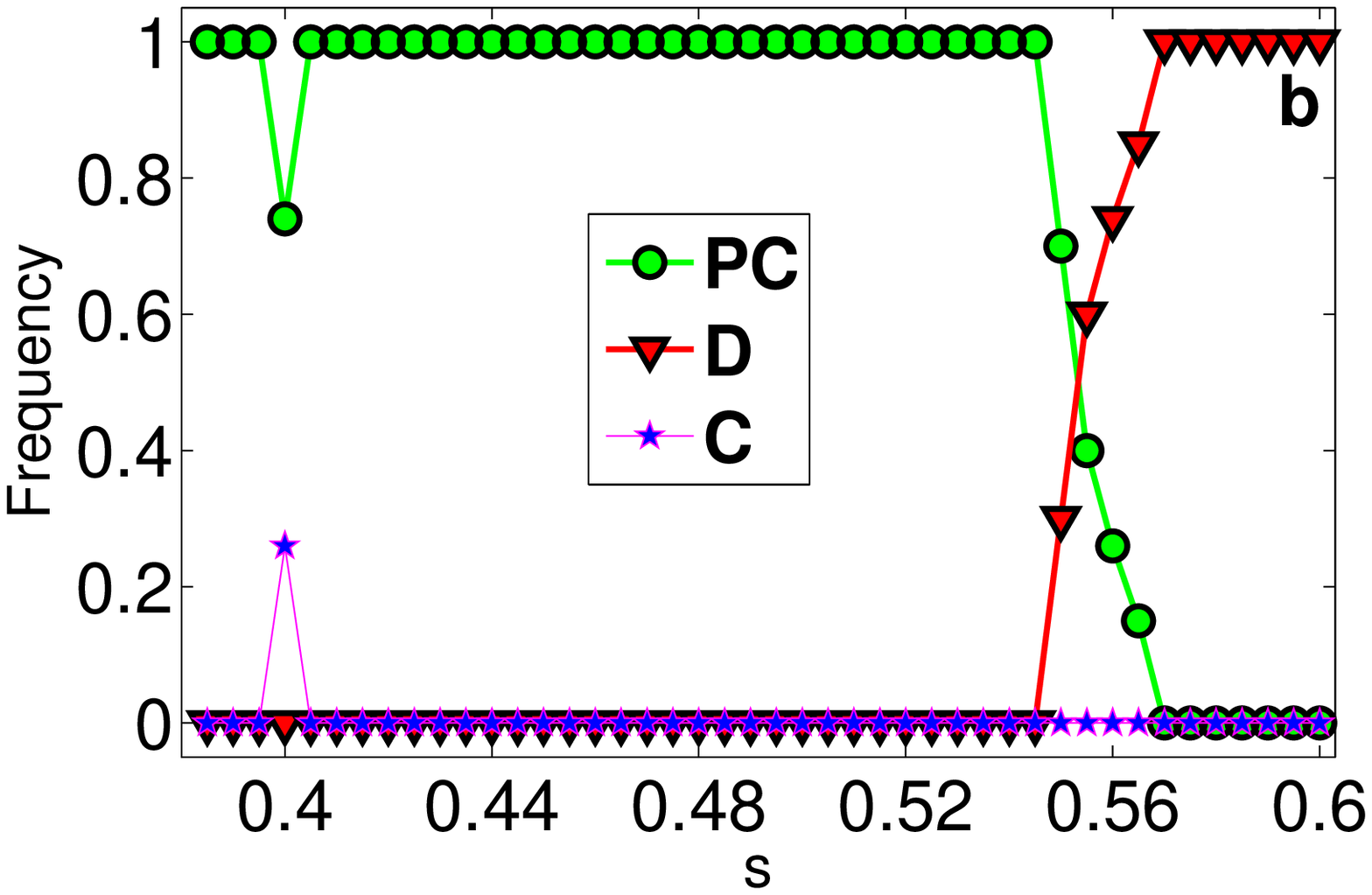} \\
  \caption{(Color online.) (a) Full $s vs d$ phase diagram obtained for $r=2$. Different phases are denoted by the symbols of strategies that survive finally in the equilibrium state. Solid lines indicate continuous transitions between different states. (b) A representative cross-section of (a) at $d=0.4$, illustrating the frequencies of $C, D$ and $PC$ in dependence on $s$. \label{fig5}}
\end{figure}
\begin{figure}
\centering
  \includegraphics[height=0.22\textheight]{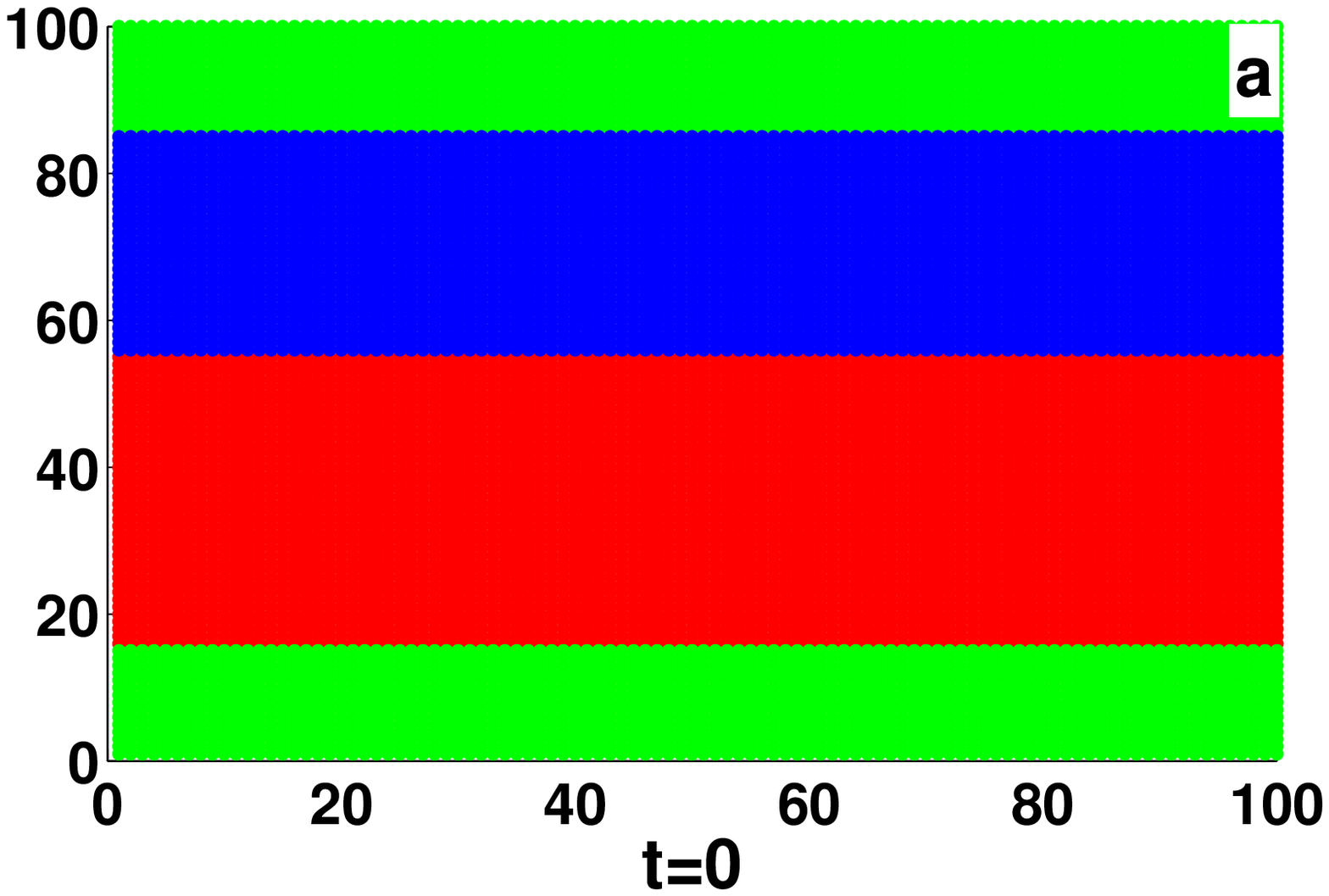} 
  \includegraphics[height=0.22\textheight]{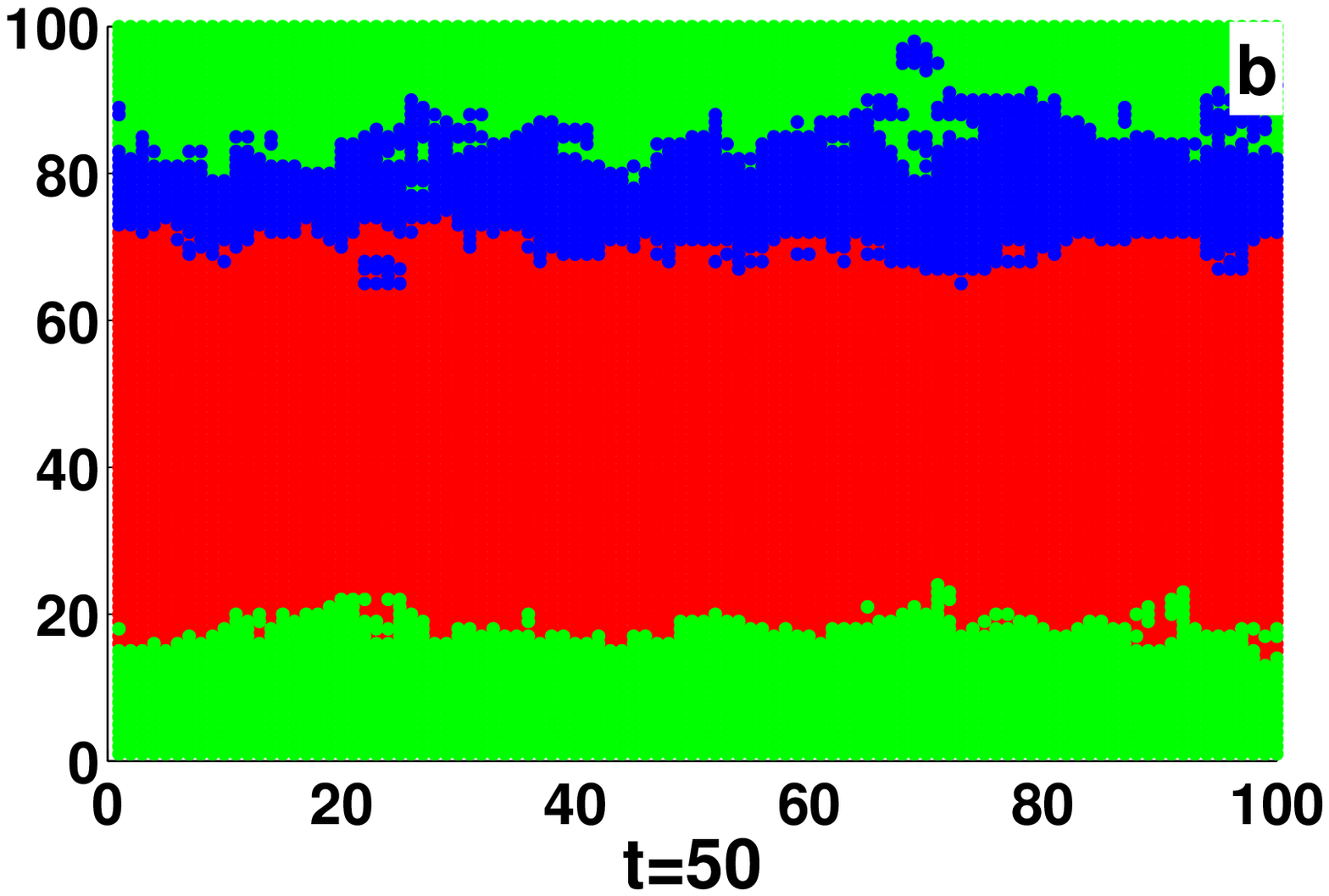} \\
  \includegraphics[height=0.22\textheight]{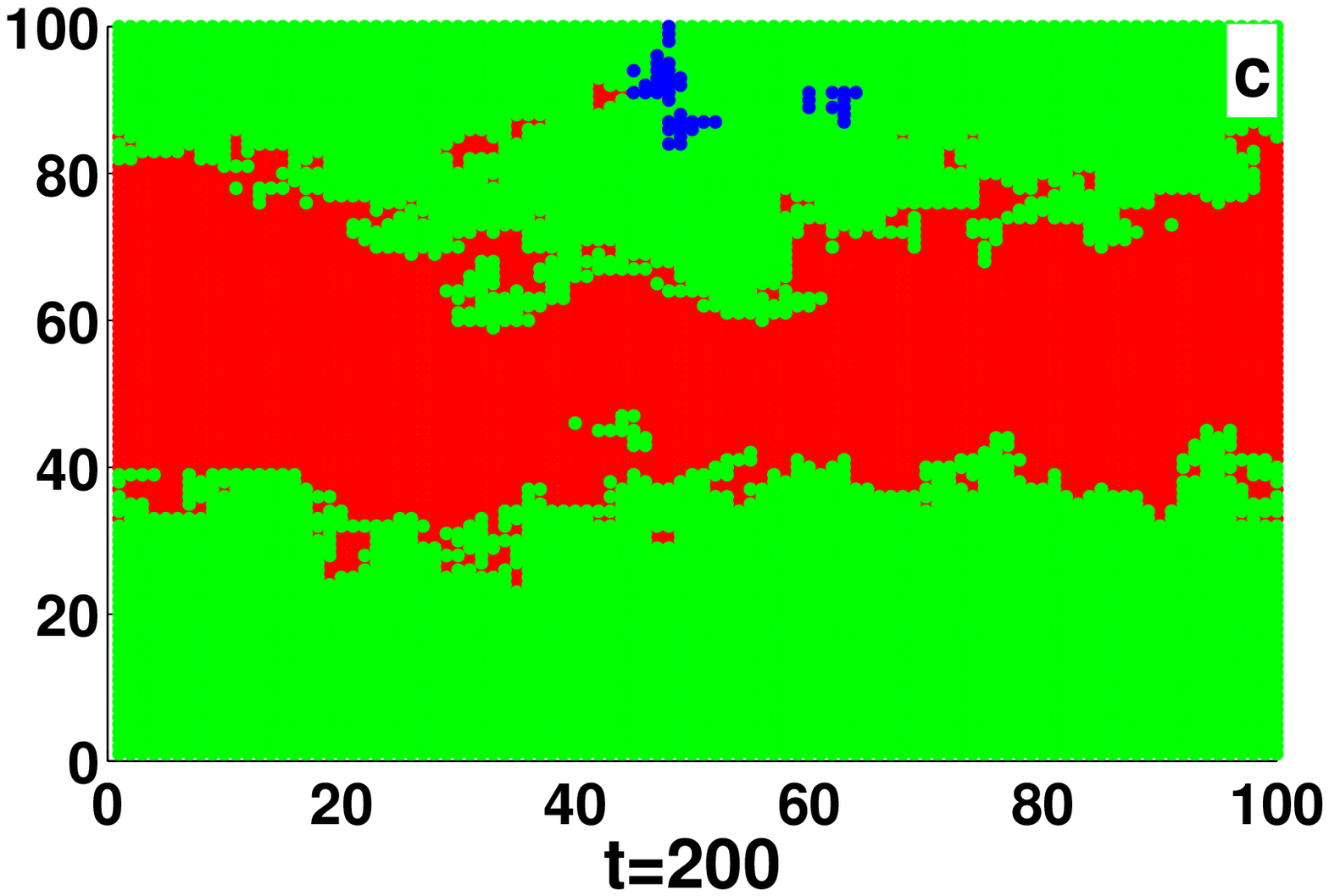}
  \includegraphics[height=0.22\textheight]{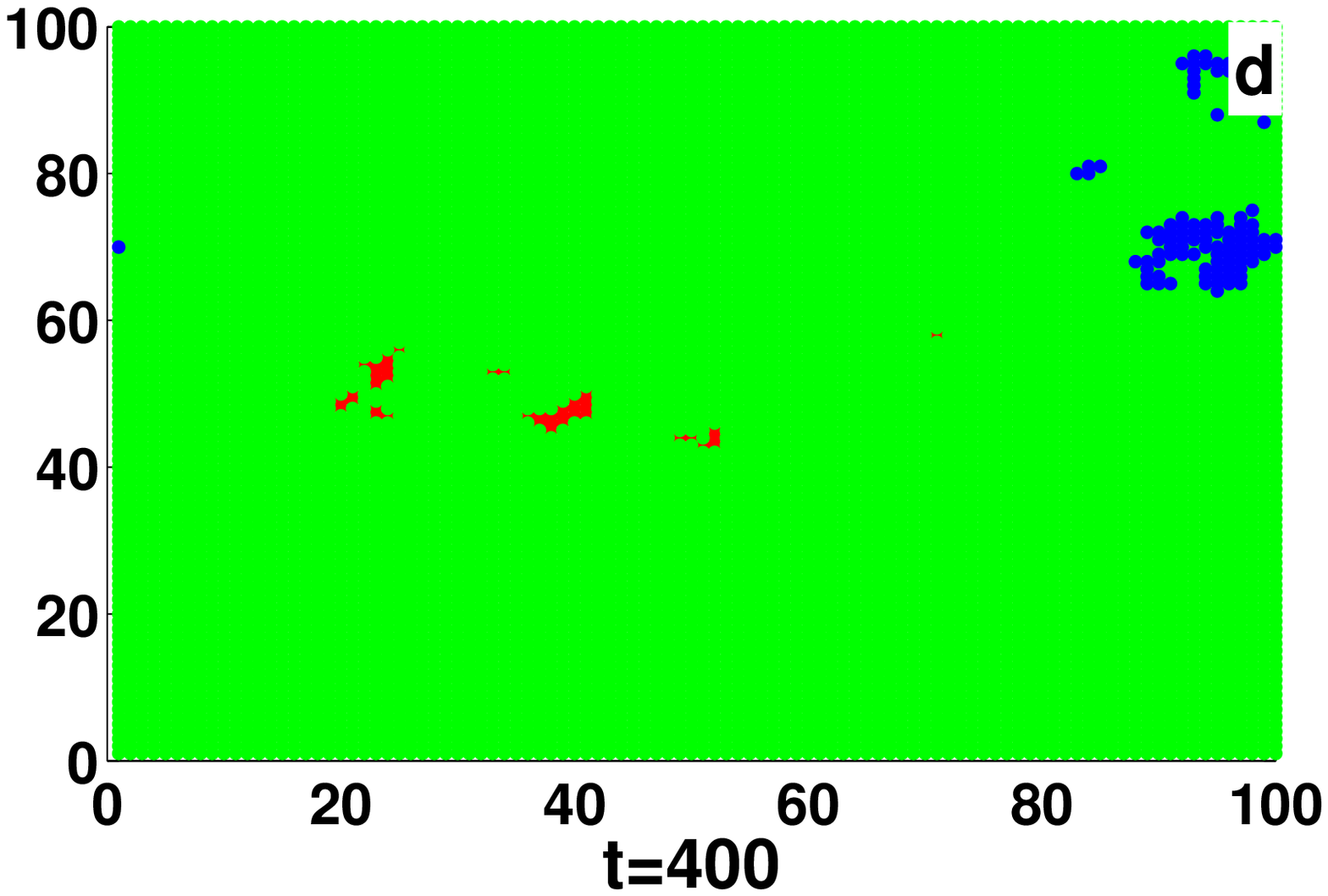}
\caption{\label{fig6}(Color online.)  Characteristic pictures of evolutionary dynamics of the three strategies on a 
100$\times$100
square lattice with specially prepared initial distribution. Colors blue, green and red show the location of cooperators ($C$), persistent cooperators ($PC$) and defectors ($D$), respectively. The snapshots were taken at (a) t=0, (b) t=50, (c) t=200 and (d) t=400 full MCs, respectively.
The parameter values were
$r=2.0$, $s=0.5$, and $d=0.4$.
}
\end{figure}

\par
We begin with $r=2$. Fig.~\ref{fig5}a depicts the full $s-d$ phase diagram, which is partitioned into several regions. From left to right,
we observe the coexisting region of $PC$ and $C$ first. In this region, only a small fraction of the total benefit is shared, so there is no space for the survival of defectors due to their low income. While for intermediate $s$,
the $D$-players obtain more of the total public goods than ever,
the persistent cooperators have to compete with the defectors for survival and retrieve their income participated by the defectors. Owing to their higher benefit, the $PC$-players perform better than the defectors.
 Therefore, the $PC$-players govern the whole population. With increasing $s$,
the $PC$-players fail to crowd out the defectors. The pure cooperators are more impressionable to be exploited by the defectors than the persistent cooperators because of their lower gain and small synergic factor $r$, they give way to the defectors. So there comes a very narrow territory for the coexistence of $PC$ and $D$. Finally, still due to the lower $r$ and that more than half of the total payoff is shared among all the participators irrespective to their strategies, it is not so tempting to retrieve the ($1-s$) fraction of the payoff contributed by themselves comparing to the second cost degree.  The $PC$-players lose their superiority in the competition. Hence all players transform their strategies and turn to be “free riders”.
\par
A typical cross-reference of Fig.~\ref{fig5}a is shown in Fig.~\ref{fig5}b at intermediate second cost degree $d=0.4$, which illustrates the frequencies of $C,$ $D$ and $PC$ in dependence on $s$.
If $s \leq 0.4$,
the suppression of treacherous actions is so powerful that it enables not only the reproduction of persistent cooperators but also the compatibility of traditional cooperators. 
It is necessary to point out that the $PC$-players get the same harvest as $C$-players
in the absence of defectors, and the two types of cooperating individuals imitate each
other with probability 0.5, now the model can be viewed as a voter model
\cite{liggett85}. 
So, when $s$ is less than the critical value as
described above, after the defectors are driven out, the whole system either falls in a
homogenous state of $PC (C)$ or a mixed state of $PC$ and $C$ occasionally according to
the initial state which is in accordance with the phase diagrams.
While $0.4<s<0.55$, the $PC$ individuals occupy the population and the frequency of $C$ becomes 0 due to the exploit by the defectors. If $s$ hits the threshold value 0.555, defectors appear, and the frequency of $D$ increases from 0 to 1 monotonously. Though the $PC$-players remain alive, the amount of them reduces sharply to zero during a short time. Each individual then choses to be a defector if $s>0.56$. This can be verified by analyzing the snapshots of strategies' distribution in Fig.~\ref{fig5}. We use a specially prepared initial state similar to that in 
\cite{szolnoki10},
which has three distinct interfaces among the three strategies. To gain insight into the influence of $PC$ and the evolution of the three strategies over time, we may focus on the movements of these interfaces that separate the three strategies. The defectors (red) outperform the pure cooperators (blue) because of the low synergic factor $r$, they can break into the territory of the $C$-players easily. But the cooperators can survive in the vicinity of persistent cooperators, as shown in Fig.~\ref{fig5}c. Meanwhile, since $s$ is not large, the defectors gain not so much payoff as they expect, they have to imitate the strategy of $PC$-players (green) around them. Therefore, persistent cooperators cross the fence between $PC$ and $D$, and inhibit the spread of strategy $D$. The defectors finally lose their territory in the competition. While $s$ is less than the critical value and $d$ is not too large, the final stable state of the population will be the dominated by the $PC$-players.
\begin{figure}
\centering
  \includegraphics[height=0.22\textheight]{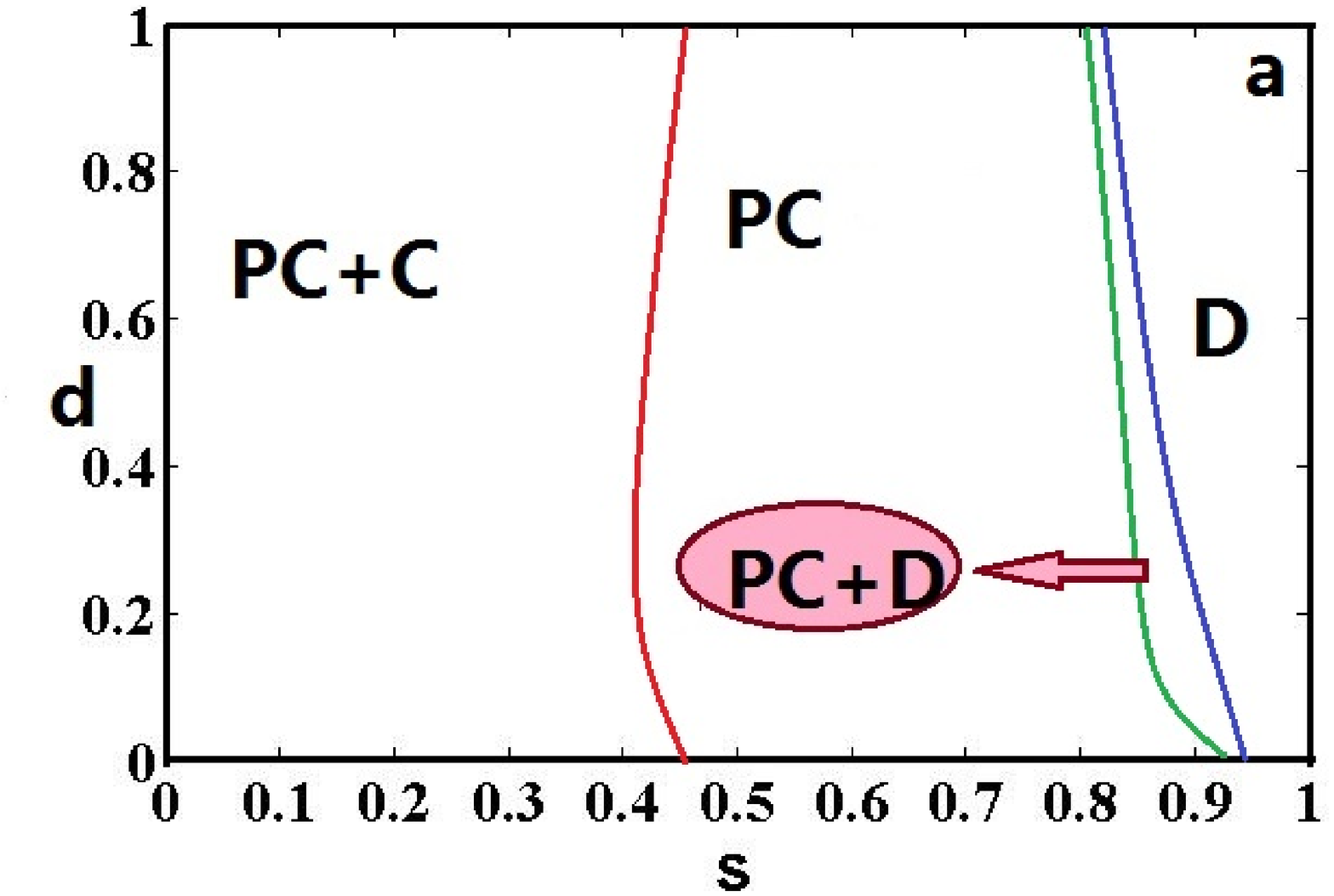} 
  \includegraphics[height=0.22\textheight]{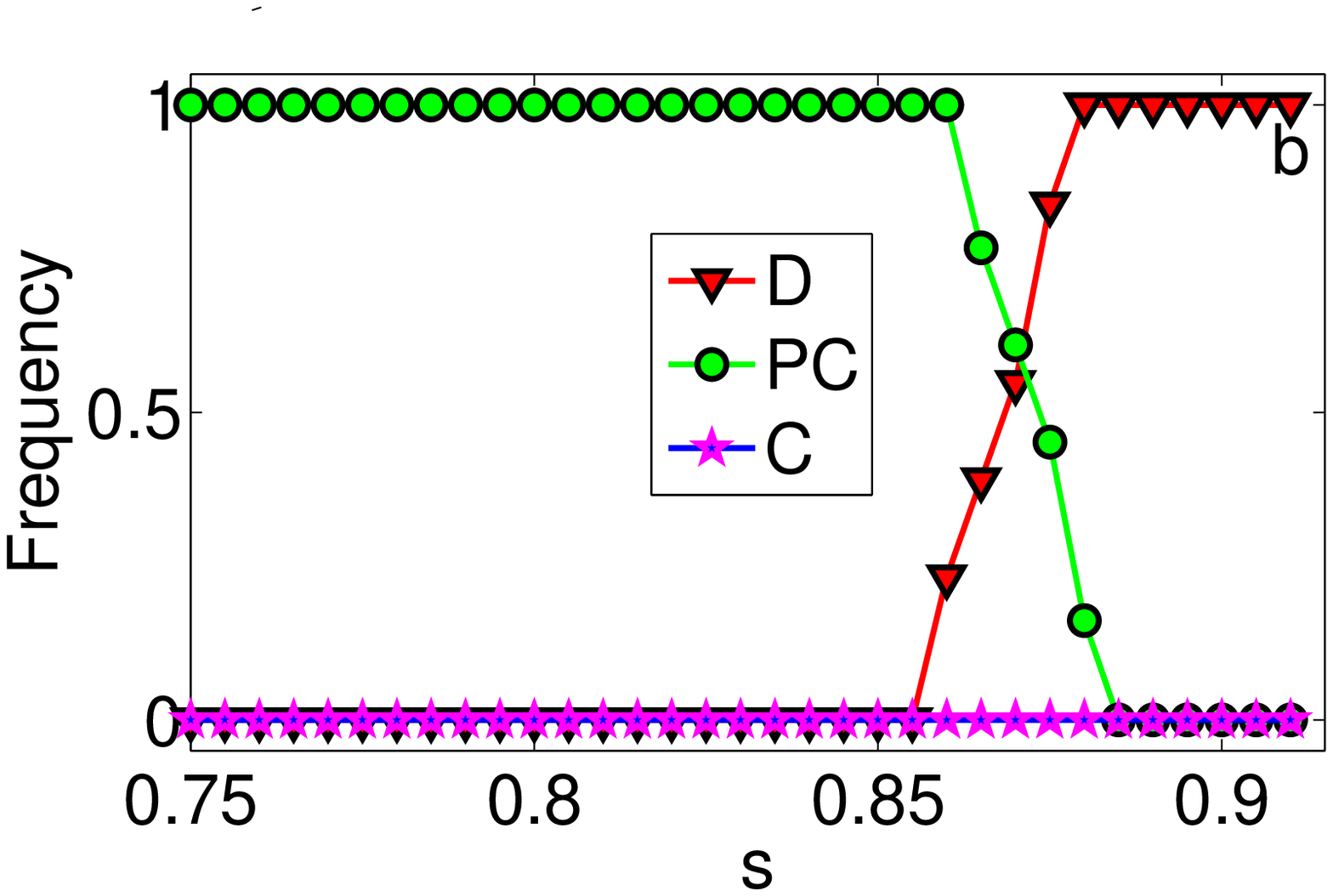} 
\caption{(Color online.)   (a) Full $s$ vs $d$ phase diagram obtained for $r=3.5$. Different phases are denoted by symbols of strategies that survive finally in the equilibrium state. Solid lines indicate continuous transitions between different states. (b) A representative cross-reference of (a) at $d=0.4$, illustrating the frequencies of $C, D$ and $PC$ depending on $s$.
\label{fig8}}
\end{figure}
\par
We proceed to discuss the evolutionary dynamics of the three strategies for relative high synergic factor $r=3.5$, at which the cooperators still be the loser in the
evolutionary process
without any supporting mechanism. Comparing to the full $s-d$ phase diagram presented in Fig.~\ref{fig5}a, it is clear that the qualitative characteristics in Fig.~\ref{fig8}a remain intact to a large degree. Similar to the observations found for $r=2$, 
 again here we detect a
wide region of the coexistence state of $PC$ and $C$. One of the most remarkable differences between the two diagrams, is of course the expansion of the coexistence region of $PC$ and $D$, owing to the spatial reciprocity \cite{nowak92}. When $s$ is quite large, the persistent cooperators disappear and leave a much narrower region for the defectors than the space in Fig.~\ref{fig5}a. This is in accordance with intuition in view of synergic factor, which promotes cooperating behavior.
\par
Fig.~\ref{fig8}b confirms the results from the quantitative aspect. 
 In our simulation, if $s$ is less than
0.855,
 which is much larger than that for $r=2$ as well as the critical value in well-mixed population and one-dimensional lattice-structured population, $PC$-players are conquerors in most time. When $s$ increases and exceeds the critical value, the frequency of $PC$ falls down to 0 as sharply as depicted in 
Fig.~\ref{fig5}b. It follows that if $s$ is larger than the threshold, the $PC$-players have the fate to be driven out 
in the competition
sooner or later for any $r<3.79$.

\section{Discussion and Summary \label{sec5}}
Mankind which engages in various complex games of cooperation and defection 
 achieves highest forms of cooperation
among all life forms on the earth \cite{nowak06}. Many mechanisms were proposed to discover the mystery why natural selection is in favor of cooperative behavior, including the new model proposed by Liu and Guo \cite{liu10}, in which the total resulted payoff is shared in a different way from the normal PGG, that is, only a proportion of the total profit is shared by all the players irrelevant to their individual contribution. They introduced a new strategy named persistent cooperation in well-mixed populations. A persistent cooperator is a contributor who is willing to pay a second cost to retrieve the remaining portion of the payoff contributed by themselves.
\par
In this paper, we focused on investigating the impact of persistent cooperating on the evolution of cooperation in 
spatial public goods games. 
The evolutionary dynamics of the structured populations consisting of three types of competing players (pure cooperators, defectors and persistent cooperators) are revealed by theoretical analysis and numerical simulations. In particular, we derive the approximate expressions of fixation probabilities for strategies on one-dimensional lattice. By using one-dimensional lattice and square lattice as the basic interacting network, the spatial patterns of cooperation are discussed by means of the phase diagrams of stationary states of strategies depending on different combinations of distribution fraction of the total payoff and the second cost degree. The results reveal many relevant differences compared with the conclusions derived in well-mixed populations. Such as, the coexistence of $D$ and $PC$ never appears in well-mixed populations but comes up in lattice-structured populations. The evolutionary time for $PC$ to reach the equilibrium state in lattice-structured populations is much shorter than that in well-mixed populations. Furthermore, the ranges of parameters for $PC$ dominating the whole lattice-structured populations are larger than that for well-mixed populations. Our results also show that persistent cooperation promotes the evolution of cooperation for low synergic factor $r$ effectively, though defection takes over a considerably large territory in full phase diagram of $s$ versus $d$. For relative high synergic factor, cooperators can survive attribute to network reciprocity. In addition, comparing the fixation probability of a $PC$-mutant with that of a $D$-mutant on one-dimensional lattice provides further understanding for persistent cooperation in structured populations. For example, in some rigorous situations the difficulty in invasion and fixation of cooperation in a population consisting of non-cooperating individuals in well-mixed populations can be solved in the lattice-structured populations.

It is also interesting to compare the persistent cooperation model studied here
with punishment models that were introduced in the context of PGGs
to eliminate free riding (defection) and falilitate cooperation
\cite{sigmund01,sigmund07}.
Whether cooperation is facilitated, however may depend --
besides network structure \cite{perc13} --
on parameters like the degree of rationality
\cite{wang10}, and, in general, on the implementation of punishment in the game.
Punishment can be introduced in different forms, namely pool punishment
\cite{szolnoki11}
and peer punishment
\cite{helbing10pre,helbing10plos}.
In comparison between
punishing cooperators and punishing defectors
Helbing et al.\ 
\cite{helbing10njp}
 showed
 that a fixed and finite
interacting neighborhood can resolve the dilemma of second-order free-riding by
separating punishing cooperators form pure cooperators. 
Letting pool and peer punishment
 compete in the game, 
Szolnoki at al.\ showed that peer punishers outperform
the pool punishers and control the system in large segments of parameters
\cite{szolnoki11b}.
Punishment mechanisms may also be applied to 
 human behavior in monetary systems:
Recently, Wang et al.\ studied
a model of tax system as an approach of punishment 
\cite{wang14}.
Overall, all punishment strategies in the previous
models are costly. 
However, the purpose of persistent cooperators in our model is
not only punishing defectors but also retrieving their deserved payoff.
One main difference
between our model and the previous ones is that the payoff of $PC$-strategy is not
always less than the pure cooperators. i.e. the benefit of $PC$-players is larger than that
of cooperators if
$(1-s)r-d>0$.
By this way, second-order free-riding problem can be
avoided in appropriate range of $s$ and $d$ for certain $r$ without incurring personal
benefit.

In conclusion, new interesting phenomena occur 
in a dynamics where agents play the PGG with persistent cooperation in
 lattice-structured populations. The existence of persistent cooperators greatly suppresses the spreading of defectors under more relaxed conditions in structured populations than that in well-mixed population. Furthermore, our results are consistent 
not only with
the conclusion that the population structures facilitate the evolution of cooperation in most situations, but also the common social experience that the higher the fraction in sharing the total benefit equally the less active
humans are 
in cooperating.

\par
This work was supported by the National NSF (61374183, 10971097, 51472117) of China, 973 program (2013CB932604, 2012CB933403), the Research Fund of State Key Laboratory of Mechanics and Control of Mechanical Structures (0414K01) and PAPD.



%
\appendix
\section{Game payoffs of steps 1 and 2}
The overall payoff 
per game round 
is divided in two stages: in the first stage, only a fraction $s (0<s<1)$ of the resulted benefit is shared equally among $n+1$ participators irrespective of their strategies. Every cooperator obtains
\begin{equation}
\nonumber
P^1_{C,n_{PC},n_C}=\frac{sr(n_{PC}+n_C+1)}{n+1}-1
\end{equation}\\
and a defector has
\begin{equation}
\nonumber
P^1_{D,n_{PC},n_C}=\frac{sr(n_{PC}+n_C)}{n+1}
\end{equation}\\
while each persistent cooperator receives
\begin{equation}
\nonumber 
P^1_{{PC},n_{PC},n_C}=\frac{sr(n_{PC}+n_C+1)}{n+1}-1
\end{equation}
\par
In the second stage, the remaining proportion of the total income $(1-s)r(n_{PC}+n_C)$  is divided into two parts: 
One part  $(1-s)rn_{PC}$ contributed by the persistent cooperators will be retrieved by themselves with persistent efforts at a second personal cost $dn_D/(n+1)$, where  $n_D/(n+1)$  is the proportion of defectors and $d>0$. Thus in this stage each persistent cooperator gets the payoff $(1-s)r-dn_D/(n+1)$. However the pure cooperators are unwilling to bear additional cost to take back this part of deserved payback $(1-s)rn_C$ , so this part of payoff is again but not equally shared among cooperators and defectors. Each defector gets $(1-s)rn_C/(n+1)$, a cooperator reaps $[(1-s)rn_C-(1-s)rn_Cn_D/(n+1)]/n_C$. Here we assume that the $PC$-players are so friendly and generous that they have no intension to share the $(1-s)$ part of the payoff contributed by the cooperators. Hence, in this stage, every cooperator earns
\begin{equation}\label{eqA4}
P^2_{C,n_{PC},n_C}=(1-s)r-\frac{(1-s)rn_D}{n+1}
\end{equation}\\[-4ex]
and a defector has
\begin{equation}\label{eqA5}
P^2_{D,n_{PC},n_C}=\frac{(1-s)rn_C}{n+1}
\end{equation}\\
while each persistent cooperator receives
\begin{equation}\label{eqA6}
P^2_{{PC},n_{PC},n_C}=(1-s)r-\frac{dn_D}{n+1}
\end{equation}
\par
 Therefore, the total payoff for each player engaged in the game are:
\begin{equation}\label{eqA7}
\begin{split}
P_{C,n_{PC},n_C}&=P^1_{C,n_{PC},n_C}+P^2_{C,n_{PC},n_C}\\
&=\frac{r(n_{PC}+n_C+1)}{n+1}-1
\end{split}
\end{equation}\\[-4ex]
and a defector has
\begin{equation}\label{eqA8}
\begin{split}
P_{D,n_{PC},n_C}&=P^1_{D,n_{PC},n_C}+P^2_{D,n_{PC},n_C}\\
&=\frac{rn_C+srn_{PC}}{n+1}
\end{split}
\end{equation}\\
while each persistent cooperator receives
\begin{equation}\label{eqA9}
\begin{split}
P_{{PC},n_{PC},n_C}&=P^1_{{PC},n_{PC},n_C}+P^2_{{PC},n_{PC},n_C}\\
&=P_{C,n_{PC},n_C}+\frac{[(1-s)r-d]n_D}{n+1}
\end{split}
\end{equation}
\par

\end{document}